\documentclass[reviewcopy]{elsarticle}

\usepackage[reviewcopy]{adndt}
\usepackage{longtable}

\usepackage{amsmath}
\usepackage{amssymb}

\biboptions{square,sort&compress}
\bibpunct[]{[}{]}{,}{n}{}{;}
\begin{document}

\begin{frontmatter}

\journal{Atomic Data and Nuclear Data Tables}

\title{Ionization probabilities of Ne, Ar, Kr,
and Xe by proton impact for different initial states and impact
energies}

\author[One,Two]{J. E. Miraglia}
\author[One,Two]{C. C. Montanari \corref{cor2}} \ead{mclaudia@iafe.uba.ar}

\cortext[cor2]{Corresponding author.}

\address[One]{Instituto de Astronom\'{\i}a y F\'{\i}sica del Espacio,\\ Consejo Nacional de
Investigaciones Cient\'{\i}ficas y T\'{e}cnicas and Universidad de
Buenos Aires,\\ casilla de correo 67, sucursal 28, (C1428EGA) Buenos
Aires, Argentina}

\address[Two]{Facultad de Ciencias Exactas y Naturales, Universidad de Buenos
Aires\\ Buenos Aires, Argentina}

\date{Today}

\begin{abstract}
Tables of \textit{ab-initio} total cross sections, probabilities at
zero impact parameter, and impact parameter moments are presented
concerning the ionization of Ne, Ar, Kr, and Xe by proton impact in
the energy range (0.1-10) MeV. The calculations correspond to the
continuum distorted wave eikonal initial state approximation
(CDW-EIS) for energies up to 1 MeV, and to the first Born
approximation for larger energies. The results displayed in the
tables are disaggregated for the different initial bound states,
considering all the shells for Ne and Ar, the L-M-N shells of Kr and
the M-N-O shells of Xe. Our inner-shell ionization cross sections
are compared with the available experimental data and with the
ECPSSR results.
\end{abstract}

\end{frontmatter}

\newpage

\tableofcontents
\listofDtables
\listofDfigures
\vskip5pc


\section{Introduction}

Ionization data, involving experimental and theoretical values, has
received great attention for a very long time (see for example
\cite{Miranda2014} and references therein). However, current
progress of beam characterization methods, atomic analytical
techniques such as the so-extended particle induced x-ray emission
(PIXE) \cite{PIXE,PIXE2013}, and multiple-purpose simulations for
the passage of particles through matter as the Geant4 \cite{geant4},
have aroused new interest and requirements of accurate data and
reliable predictions, and have also shown certain vacancy areas for
the theoretical development.

Reliable values for the probabilities as function of the impact
parameter are the seeds to describe the total ionization cross
sections of the different shells. But also, these probabilities are
the inputs for the multiple ionization calculations
in a multinomial combination of the impact parameter probabilities
\cite{Olson}. From the theoretical point of view, these
probabilities represent a challenge and a test of the capability of
a theory to describe wave functions and interaction potentials.
Different approaches have been employed over the years, from the
basic first Born approximation, to distorted wave methods, numerical
solution of the Schr\"odinger equation or collective response models
\cite{Spranger04,Cavalcanti02,MM1,AP,Galassi,archubi}. Moreover, in
very recent works, the ionization probabilities by proton and
antiproton impact  have been the seeds to obtain multiple ionization
cross sections of rare gases by electron and positron impact, with
reasonably good results \cite{e-rare,positron}. It is worth to note
that multiple ionization cross sections are highly dependent on the
inner-shell ionization probabilities, which contribute to the final
values through Auger-type processes \cite{MM1,e-rare}.

There are different compilations of experimental data for the total
ionization cross sections of the K-shell \cite{paul1,paul2} and the
L-shell \cite{Miranda2014,orlic}. One of the most employed models
for K and L-shell ionization cross sections is the ECPSSR by Brandt
and Lapicki \cite{brandt-lapicki,lapicki02}, of high efficiency and
the usual input in PIXE codes \cite{lapickiNIMB14}. Instead,
reliable values of M-shell ionization are scarce \cite{pajek,mitra}.
This is due to the complexity of the M-X-ray spectra because of the
existence of five subshells \cite{Mukoyama15}. It is usual that the
experimental data are presented as X-ray production cross sections.
The conversion of the X-ray production cross sections to the total
ionization cross sections depends on multiple ionization parameters
and fluorescence yields. Different possible values are discussed
\cite{miranda02,segui09}; however, the current state of the
experimental techniques \cite{campbell,campbell2} and the advance in
X-ray spectrometers represent an improvement on this respect
\cite{Miranda2014}.

The goal of the present tabulation is to make available
\textit{ab-initio} CDW-EIS and first Born  approximation results for
proton impact ionization of the heaviest rare gases, for the
different subshells. Presents results are calculated as in
\cite{miraglia08,miraglia09,miraglia10}, by rigorously solving the
radial Schr\"{o}dinger equation for different angular momenta for
both the initial bound and the final continuum states. Thus, we can
assure the proper description of the continuum wave function and its
mathematical orthogonality to the bound state. These values have
already been tested in total \cite{miraglia08} and differential
\cite{miraglia09} ionization cross sections. Also the probabilities
as function of the impact parameter have been employed in multiple
ionization calculations \cite{AP,e-rare,positron,sigaud13} with good
agreement with the experimental data, even for sextuple ionization
of Kr (Xe), where L-shell (M-shell) contribution is decisive.

In the following sections we make available these CDW-EIS values for
proton impact energies 0.1-1 MeV, and also the first Born
approximation results for proton energies 1-10 MeV, considering the
ionization of different subshells: Ne (1s, 2s, 2p), Ar (1s,..., 3p),
Kr (2s,..., 4p) and Xe (3s,..., 5p). We display total ionization
cross sections, probabilities at zero impact parameter, and impact
parameter moments of order 1 and $-1$. We also include the
comparison of our inner-shell ionization cross sections with the
available experimental data, and with the ECPSSR values
\cite{brandt-lapicki}, for the K-shell of Ne, K and L-shells of Ar,
L and M-shells of Kr and M and N-shells of Xe. Atomic units are used
throughout this work, except when specifically mentioned.

\section{Total ionization cross sections}\label{sec2}

The total ionization cross section of an electron initially in the
$nlm$ state, due to the interaction with a heavy projectile of
charge $Z_{P}$ (in this work $Z_{P}=1$, proton impact) and impact
velocity $v$, is given by the four-dimension integral

\begin{equation}
\sigma _{nlm}=\frac{(2\pi )^{2}}{v^{2}}\int d\overrightarrow{k}\int d%
\overrightarrow{\eta }\ \left\vert
T_{\overrightarrow{k},nlm}(\overrightarrow{\eta })\right\vert ^{2}
\label{10}
\end{equation}%
where $T_{\overrightarrow{k},nlm}(\overrightarrow{\eta })$ is the
transition matrix as a function of the momentum transferred
$\overrightarrow{\eta }$ perpendicular to the incident velocity
$\vec{v}$,  and $\overrightarrow{k}$ is the momentum of the emitted
electron. For heavy projectiles such as protons, the integration
over $\overrightarrow{\eta }$ extends to infinity.

If we are interested in high energy collisions, we can resort to the
first Born approximation. This is a perturbative method valid at
large impact velocities and low projectile charges, with the initial
and final wave functions being the unperturbed ones. The first Born
transition matrix element is given by
\begin{equation}
T_{\overrightarrow{k},nlm}^{Born}(\overrightarrow{\eta })=%
\frac{1}{(2\pi )^{3/2}}\widetilde{V}_{P}(p)\;\int d\overrightarrow{r}%
\;\varphi _{\overrightarrow{k}}^{\ast }(r)\ \exp (i\;\overrightarrow{p}.%
\overrightarrow{r})\ \varphi _{nlm}(r).  \label{20}
\end{equation}%
Here $\varphi _{_{\overrightarrow{k}}}$ ($\varphi _{nlm}$) is the
final (initial) continuum (bound) eigenfunction of the target
hamiltonian; $\widetilde{V}_{P}(p)=-\sqrt{2/\pi }\ Z_{P}/p^{2}$ is
the Fourier transform of the projectile-electron Coulomb potential,
and $\overrightarrow{p}=
\overrightarrow{K}_{i}-\overrightarrow{K}_{f}$ is the momentum
transferred, with $\overrightarrow{K}_{i}$
($\overrightarrow{K}_{f}$) being the initial (final) projectile
momentum.  The momentum transferred can also be expressed as
$\overrightarrow{p}=(\overrightarrow{\eta },p_{m})$, with $p_{m}$
being the minimum momentum transfer along the direction of
$\vec{v}$, $p_{m}=(\varepsilon _{f}-\varepsilon _{i})/v$, and
$\varepsilon _{i}$ ($\varepsilon _{f}$) the initial (final) electron
energy.

If we are interested in the intermediate energy regime, i.e. proton
energies smaller than 1 MeV, we have to improve the calculation of
the transition matrix. To that end we resort to the rigorous
calculations using the CDW-EIS approximation
\cite{miraglia08,miraglia09}. This model, initially proposed by
Crothers and McCann \cite{Crothers}, is one of the most reliable
approximations to deal with calculations of ionization probabilities
in the intermediate to high energy regime \cite{CDWEIS1,CDWEIS2}.
Within the CDW-EIS approach the distorted wave functions include the
projectile distortion in the initial and final channels. The
T-matrix element reads

\begin{equation}
T_{\overrightarrow{k},nlm}^{CDW-EIS}(\overrightarrow{\eta
})=-\frac{1}{(2\pi )^{3/2}}\
\overrightarrow{W}(\overrightarrow{p})\cdot
\overrightarrow{G}_{\overrightarrow{k},nlm}(\overrightarrow{p}),
\label{30}
\end{equation}
with
\begin{equation}
\overrightarrow{G}_{\overrightarrow{k},nlm}(\overrightarrow{p})=\int d%
\overrightarrow{r}\;\left[ \overrightarrow{\nabla }\varphi
_{\overrightarrow{ k}}(\overrightarrow{r})\right] ^{\ast
}e^{i\;\overrightarrow{p}\cdot
\overrightarrow{r}}\ \varphi _{nlm}(\overrightarrow{r}),  \label{40} 
\end{equation}

\begin{equation}
\overrightarrow{W}(\overrightarrow{p})=N_{\xi }\int
\frac{d\overrightarrow{ r}}{(2\pi )^{3/2}}\
E_{\overrightarrow{v}}^{+}(\overrightarrow{r})\ e^{-i\;
\overrightarrow{p}\cdot \overrightarrow{r}}\ \overrightarrow{\nabla
}F_{\overrightarrow{v}}^{-^{\ast }}(\overrightarrow{r}), \label{50}
\end{equation}

\begin{equation}
E_{\overrightarrow{v}}^{+}(\overrightarrow{r})=\exp \left[ -i\xi \ln (v\ r+%
\overrightarrow{v}\cdot \overrightarrow{r})\right] ,  \label{60} 
\end{equation}

\begin{equation}
F_{\overrightarrow{v}}^{-}(\overrightarrow{r})=\ _{1}F_{1}(-i\xi ,1,-iv\ r-i%
\overrightarrow{v}\cdot \overrightarrow{r}),  \label{70}
\end{equation}
and
\begin{equation}N_{\xi }=\exp (\xi /2)\Gamma (1+i\xi ),
\end{equation}
where $\xi =Z_{P}/v$, and $_{1}F_{1}$ is the confluent
hypergeometric. In our case, both $\varphi _{_{nlm}}$ and $\varphi
_{_{\overrightarrow{k}}}$ are numerical solutions of the same
Hamiltonian, therefore fully orthogonal, then the transition matrix
$T_{if}^{CDW-EIS}$ does not display prior-post discrepancies.

The initial bound and final continuum radial wave functions were
obtained by using the RADIALF code developed by Salvat and
co-workers \cite{salvat}. The number of pivots used to solve the
Schr\"{o}dinger equation rounds a few thousands of points, depending
on the number of oscillations of the continuum. The radial
integration was performed using the cubic spline technique. The
number of angular momenta considered, $l_{\max }$, varied between 8,
at very low ejected-electron energies, up to 28, at the largest
energies considered. The same number of azimuth angles were required
to obtain the fourfold differential cross section 
(on $\overrightarrow{k}$ and $\overrightarrow{\eta}$). Each total
cross section $\sigma _{nlm}$ in Eq. \eqnref{10} was calculated
using 35 to 199 momentum transfer values of $\eta$ to determine a
doubly differential cross section, 28 fixed electron angles and
around 40 electron energies ($E=k^2/2$), depending on the projectile
impact energy.

In tables \ref{table1}-\ref{table4} we report our results for proton
impact total ionization cross sections of the four heaviest rare
gases: Ne, Ar, Kr, and Xe. CDW-EIS values are displayed for impact
energies 0.1-1 MeV, and the first Born approximation for 1-10 MeV. A
comparison between them is possible at 1 MeV. In all these cases we
show separately the contributions of the different $nlm$ initial
states, raging from the valence shell to two deeper shells.

\section{Inner-shell ionization}

\begin{figure}[ht!]
\centering
\includegraphics[width=.82\linewidth,height=4.2in]{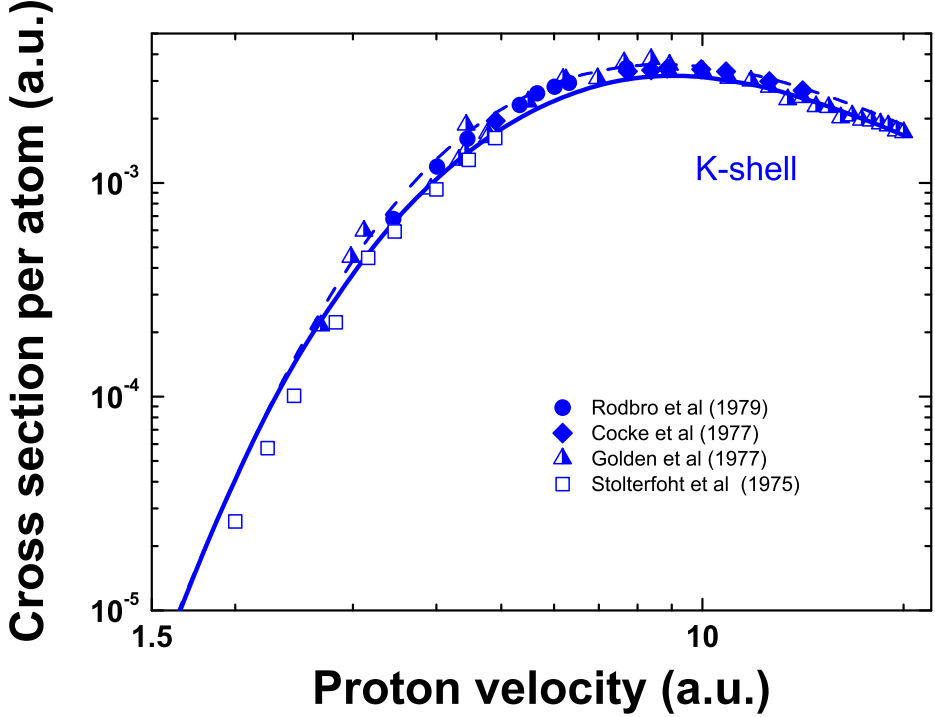}
\caption{K-shell ionization cross section of Ne by proton impact.
Curves: solid line, present results displayed in table \ref{table1};
dashed line, ECPSSR \cite{brandt-lapicki,isics}. Symbols:
experimental data by Rodbro \textit{et al} \cite{rodbro79}, Cocke
\textit{et al} \cite{cocke77}, Golden \textit{et al} \cite{golden},
and Stolterfoht \textit{et al} \cite{stolt75}. } \label{fig1}
\end{figure}


The total ionization cross sections are determined mainly by the
outer target shells. The CDW-EIS and first Born approximation values
presented here proved to be effective in the description of these
total cross sections, as shown in \cite{miraglia08}. On the other
hand, the cross sections for highly ionized targets (i.e. triple
ionization of Ar, quintuple ionization of Kr, sextuple ionization of
Xe) strongly depend on inner-shell contributions due to Auger
cascade processes \cite{sigaud13,e-rare}.

\begin{figure}[ht!]
\centering
\includegraphics[width=.82\linewidth,height=4.2in]{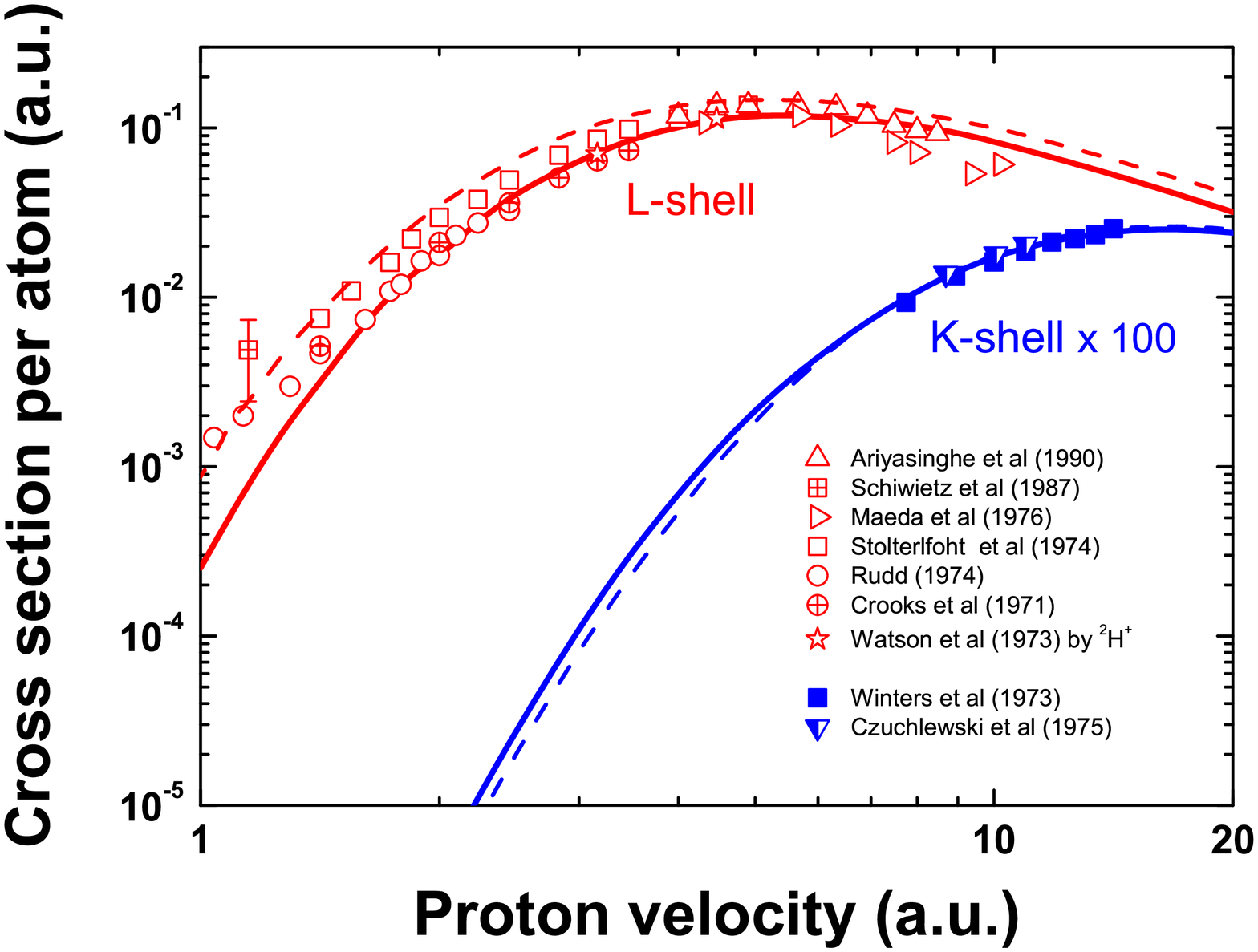}
\caption{K and L-shell ionization cross sections of Ar by proton
impact. Curves: solid line, present results displayed in table
\ref{table2}; dashed line, ECPSSR \cite{brandt-lapicki,isics}.
Symbols: K-shell experimental data by Winters \textit{et al}
\cite{winters73}, and by Czuchlewski \textit{et al} \cite{czu};
L-shell experimental data by Ariyasinghe \textit{et al} \cite{ari},
Schiwietz \textit{et al} \cite{schiwietz}, Maeda \textit{et al}
\cite{maeda}, Stolterfoht \textit{et al} \cite{stolt74}, Rudd
\cite{rudd}, Crooks and Rudd \cite{crooks}, and Watson and Toburen
\cite{watson} (deuteron impact).} \label{fig2}
\end{figure}

\begin{figure}[ht!]
\centering
\includegraphics[width=.82\linewidth,height=4.2in]{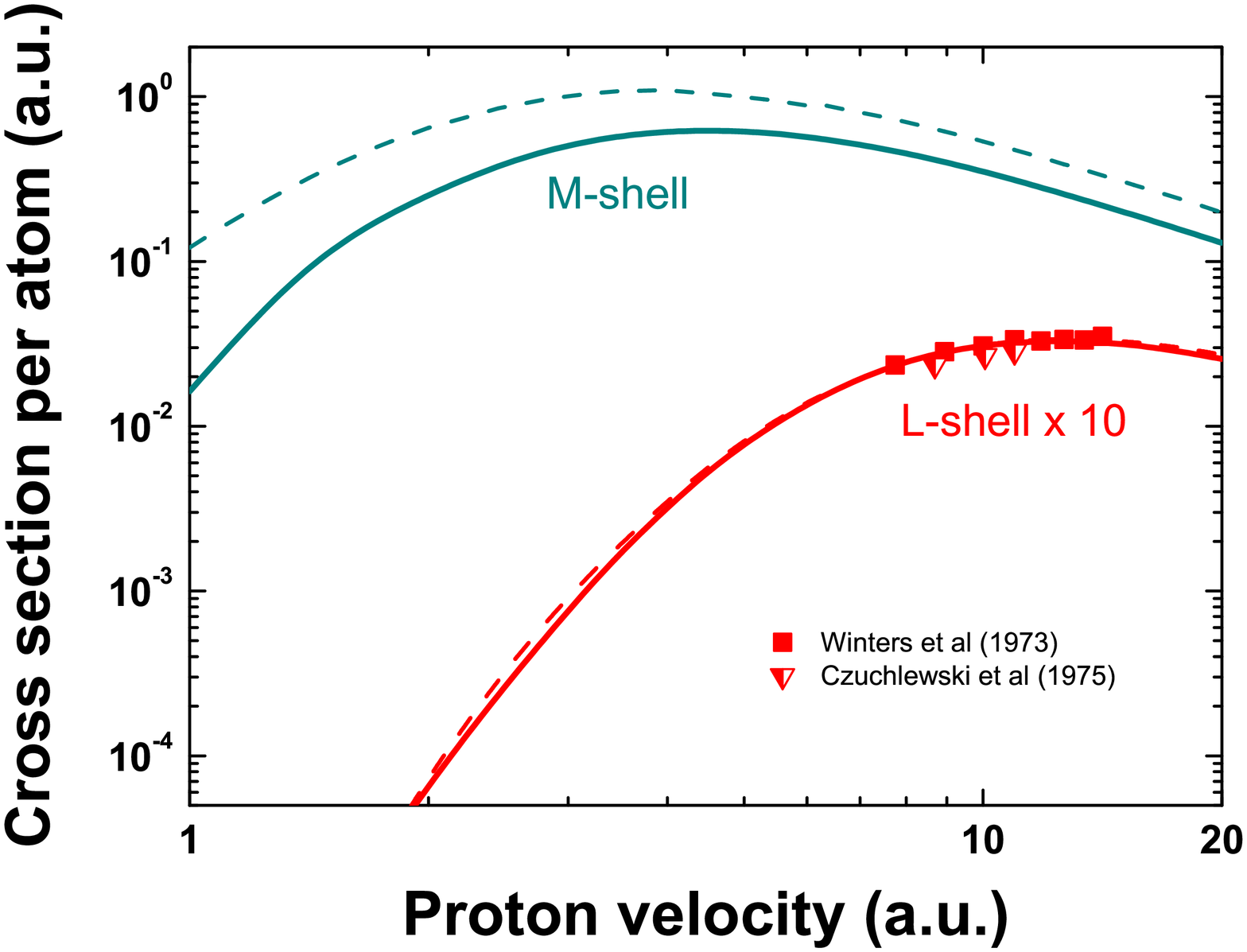}
\caption{L and M-shell ionization cross sections of Kr by proton
impact. Curves: solid line, present results displayed in table
\ref{table3}; dashed line, ECPSSR. Symbols: experimental data for
L-shell ionization by Winters \textit{et al} \cite{winters73}, and
by Czuchlewski \textit{et al} \cite{czu}.} \label{fig3}
\end{figure}
\begin{figure}[ht!]
\centering
\includegraphics[width=.82\linewidth,height=4.2in]{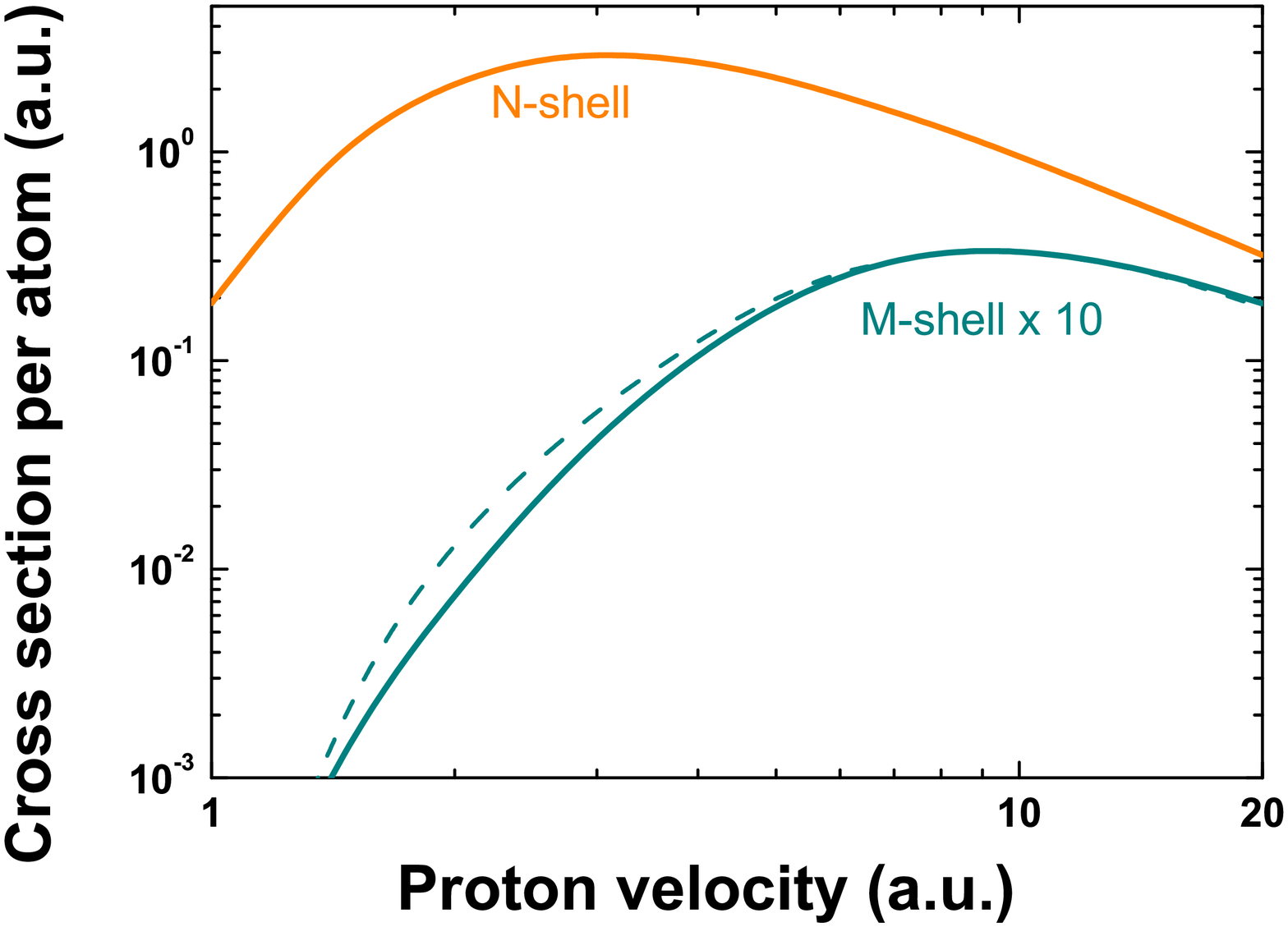}
\caption{M and N-shell ionization cross sections of Xe by proton
impact. Curves: solid line, present results displayed in table
\ref{table4}; dashed line, ECPSSR. } \label{fig4}
\end{figure}

In order to examine further the theoretical values displayed in
tables \ref{table1}-\ref{table4}, in figures \ref{fig1}-\ref{fig4}
we display these values in comparison with the experimental data
available (only for the K and L-shells) and with the ECPSSR values
obtained using the ISICS11 code by Cipolla \cite{isics} (we take
note of the comments by Smit and Lapicki \cite{lapicki14} about this
code). To our knowledge, no measurements have been reported for the
M-shell ionization of Kr and Xe, or the N-shell ionization of Xe by
proton impact.

The K-shell ionization cross section displayed in figure \ref{fig1}
shows good agreement with the data. Note that for proton impact
energy above 700 keV, the multiple ionization of Ne is sensitive to
the K-shell ionization. It contributes directly to multiple
ionization via single ionization of one K-shell electron followed by
postcollisional electron emission via Auger processes. Similar
importance has the L-shell of Ar displayed in figure \ref{fig2}. For
this shell we include in figure \ref{fig2} the proton impact data
compiled by Miranda and Lapicki \cite{Miranda2014}, and the deuteron
impact data by Watson and Toburen \cite{watson}. For the ionization
of Ar K-shell the agreement with the experiments and the ECPSSR is
very good. For L-shell ionization, our description differs from the
ECPSSR one, while the experimental data seems to be in between both
curves.

Perhaps the most interesting feature is displayed in figure
\ref{fig3} for Kr. For L-shell ionization the agreement of the
present results with the measurements by Winters \textit{et al}
\cite{winters73} and Czuchlewski \textit{et al} \cite{czu}, and with
the ECPSSR is very good. However, for M-shell ionization the
situation is different: we got a clear discrepancy with the ECPSSR
results \cite{isics} and no experimental data is available. Present
M-shell ionization probabilities of Kr have already been employed in
multiple ionization calculations with very good agreement with the
experimental data \cite{AP,sigaud13}. In triple and quadruple
ionization of Kr at high energies the M-shell ionization plays a
mayor role due to single ionization followed by Auger emission of 2
or more electrons. This is also a test of the present results
validity.

Finally, in figure \ref{fig4} we display the M and N-shell
ionization cross sections of Xe. Good agreement with ECPSSR is
observed for the M-shell ionization at high energies, and a clear
difference below 500 keV. The M-shell of Xe is very important in
quintuple and sextuple ionization above 1 MeV
\cite{sigaud13,e-rare}, which is the region where both calculations
agree well.

\section{Ionization probabilities as a function of the impact parameter}

The transition amplitude $a_{\overrightarrow{k},nlm}(%
\overrightarrow{b})$ as a function of the impact parameter $%
\overrightarrow{b}$ is defined with the help of the bi-dimensional
Fourier transform
\begin{equation}
a_{\overrightarrow{k},nlm}(\overrightarrow{b})=\int %
d\overrightarrow{\eta }\frac{\exp (i\overrightarrow{b}\cdot \overrightarrow{%
\eta })}{2\pi }\ T_{\overrightarrow{k},nlm}(\overrightarrow{%
\eta }).  \label{100}
\end{equation}%
with the probability being
$P_{\overrightarrow{k},nlm}(\overrightarrow{b})=\left\vert
a_{\overrightarrow{k},nlm}(\overrightarrow{b})\right\vert ^{2}$. To
solve (\ref{100}) we expanded
\begin{equation}
T_{\overrightarrow{k},nlm}(\overrightarrow{\eta }%
)=\sum\limits_{\mu =-M}^{M}i^{\mu }\frac{\exp (i\mu \ \varphi _{\eta })}{%
\sqrt{2\pi }}\ T_{\overrightarrow{k},nlm}^{(\mu )}(\eta ).
\label{110}
\end{equation}%
where $\overrightarrow{\eta }=\left\{ \eta ,\varphi _{\eta }\right\}
$. We integrated numerically the T-matrix elements for different
angular momentum and added them appropriately. To be consistent, the
maximum value M was
considered to be the maximum angular momenta used to solve the Schr\"{o}%
dinger equation. Special care should be taken to obtain $T_{\overrightarrow{%
k},nlm}^{(\mu )}(\eta )$ (see details in \cite{AP}). For practical purposes all these $T_{\overrightarrow{k}%
,nlm}^{(\mu )}(\eta )$ values were stored in a large table of $\
(2\times 8+1)$ to ($2\times 28+1)$ values of $\mu $, around 70
values of $\eta $, 28 electron angles $\Omega $, and between 33 to
45 values of $E$, which are available upon request.

Afterwards, equation (\ref{100}) can be written as
\begin{equation}
a_{\overrightarrow{k},nlm}(\overrightarrow{b}%
)=\sum\limits_{\mu =-M}^{M}i^{\mu }\frac{\exp (i\mu \ \varphi _{\eta })}{%
\sqrt{2\pi }}\ a_{\overrightarrow{k},nlm}^{(\mu )}(b), \label{120}
\end{equation}%
with
\begin{equation}
a_{\overrightarrow{k},nlm}^{(\mu )}(b)=i^{-\mu
}\int\limits_{0}^{\infty }d\eta\ \eta \ \ J_{\mu }(b\ \eta )\ T_{\overrightarrow{%
k},nlm}^{(\mu )}(\eta ),  \label{130}
\end{equation}%
and with $J_{\mu }(b\ \eta )$ being the cylindrical Bessel function.
Then the total ionization probability as a function of the impact
parameter is obtained after integrating in the ejection electron
space
\begin{equation}
P_{nlm}(\overrightarrow{b})=\frac{1}{2\pi }\sum\limits_{m=-M}^{M}\int d%
\overrightarrow{k}\ \left\vert a_{\overrightarrow{k},nlm}^{(\mu
)}(b)\right\vert ^{2}.  \label{140}
\end{equation}%
It is always convenient to re-calculate the total cross section as
$\sigma _{nlm}=\int d\overrightarrow{b}\
P_{nlm}(\overrightarrow{b})$ to check the procedure.

These impact parameter dependent probabilities are very important
because they are the seeds to get the different multiple ionization
ones by introducing them in the multinomial expansion
\cite{Spranger04,MM1}. If you are interested in multiple processes,
the behaviour of the probability at small impact parameters happens
to be very important. In tables \ref{table6}-\ref{table9} we display
the geometrical factor $P_{nlm}(0)$ for the four rare gases,
considering the different subshells and impact energies.

\section{Impact parameter moments}

We define the different impact parameter moments as
\begin{equation}
\left\langle b_{nlm}^{J}\right\rangle =\frac{1}{\sigma _{nlm}}\int d%
\overrightarrow{b}\ P_{nlm}(\overrightarrow{b})\left(
\frac{b}{\left\langle r_{nl}\right\rangle }\right) ^{J}.
\label{150}
\end{equation}%
To bring these moments to unity, we have normalized to $\left\langle
r_{nl}\right\rangle $, the mean radio of the initial atomic state
(see Table \ref{table5}). In Tables \ref{table10}-\ref{table17} we
display $\left\langle b_{nlm}^{1}\right\rangle $ and $\left\langle
b_{nlm}^{-1}\right\rangle ~$ for Ne, Ar, Kr and Xe and different
subshells. These are two important parameters, specially to compare
with classical trajectory Monte Carlo calculations. The classical
microcanonical ensamble, usually used to describe the initial state
velocity distribution, has a finite space dimension. It produces
P(b) which falls down abruptly with b. Thus, total cross sections
are generally well reproduced at expenses of an enhancement of
$P(0)$. A comparison with our results on these two regions: $P(0)$
and $\left\langle b_{nlm}^{1}\right\rangle$ can shed some light on
the reliability of the approach.

\section{Conclusions}

We tabulated the results of \textit{ab-initio} total cross sections,
probabilities at zero impact parameter, and two impact parameter
moments for ionization of Ne, Ar, Kr, and Xe by proton impact, using
the CDW-EIS approximation for energies up to 1 MeV and the first
Born approximation for higher energies, up to 10 MeV. We have
considered all the shells of Ne and Ar, the L-M-N shells of Kr, and
the M-N-O shells of Xe. In this way we have described the ionization
probability through four values: $\sigma _{nlm}$, $P_{nlm}(0)$,
$\left\langle b_{nlm}^{1}\right\rangle $ and $\left\langle
b_{nlm}^{-1}\right\rangle$. Imposing these four conditions to a
reasonable trial expression for $P_{nlm}(\overrightarrow{b})$ one
can describe most of the physics involved, even for multiple
ionization processes.

\ack The authors acknowledge the financial support from the
following Argentine institutions: Consejo Nacional de
Investigaciones Cient\'{\i}ficas y T\'{e}cnicas (CONICET),
Universidad de Buenos Aires, through the programme UBACyT, and
Agencia Nacional de Promoci\'{o}n Cient\'{\i}fica y Tecnol\'{o}gica
(ANPCyT). They also thank Professors Javier Miranda and Gregory
Lapicki for their interesting comments on this subject and for
kindly sharing their tables of data.



\newpage

\TableExplanation

To save space, throughout these tables the subindex $\pm n$ replaces
the  $10^{\pm n}$ factor.

\bigskip

\renewcommand{\arraystretch}{1.0}

\section*{Table 1.\label{tbl1te} Total ionization cross sections, $\sigma_{nlm}$,
of Neon K and L shells by 0.1-10 MeV protons. Atomic units are
used.}

\begin{tabular*}{0.90\textwidth}{@{}@{\extracolsep{\fill}}lp{5.5in}@{}}
$E$     & proton impact energy  in MeV \\
$nlm$   & electron initial state \\
CDW-EIS & Total cross section per electron, $\sigma _{nlm}$, in
atomic units, given by Eq. (\ref{10}), with the CDW-EIS T-matrix
element given
by Eq. (\ref{30}), for impact energies 0.1-1 MeV \\
Born    & Total cross section per electron, $\sigma _{nlm}$, in
atomic units, given by Eq. (\ref{10}), with the first Born T-matrix
element given
by Eq. (\ref{20}), for impact energies 1-10 MeV \\
\end{tabular*}

\renewcommand{\arraystretch}{1.0}
\bigskip

\section*{Table 2.\label{tbl2te} Total ionization cross sections, $\sigma_{nlm}$,
of  Argon K, L and M shells by 0.1-10 MeV protons. Atomic units are
used. Explanation as in Table 1.}

\bigskip

\section*{Table 3.\label{tbl3te}
Total ionization cross sections, $\sigma_{nlm}$, of  Krypton L, M
and N shells by 0.1-10 MeV protons. Atomic units are used.
Explanation as in Table 1.}


\bigskip

\section*{Table 4.\label{tbl4te}
Total ionization cross sections, $\sigma_{nlm}$, of Xenon M, N and O
shells by 0.1-10 MeV protons. Atomic units are used. Explanation as
in Table 1.}

\bigskip

\section*{Table 5.\label{tbl5te}
Ionization probabilities at zero impact parameter of Neon K and L
shells by 0.1-10 MeV protons.}
\begin{tabular*}{0.90\textwidth}{@{}@{\extracolsep{\fill}}lp{5.5in}@{}}
$E$ & proton impact energy  in MeV \\
$nlm$ & electron initial state \\
CDW-EIS & Ionization probability per electron at impact parameter
$b=0$, $P _{nlm}(0)$, given by Eq. (\ref{140}), calculated within
the CDW-EIS approximation as described in section \ref{sec2}, for impact energies 0.1-1 MeV \\
Born & Ionization probability per electron at impact parameter
$b=0$, $P _{nlm}(0)$, given by Eq. (\ref{140}), calculated within
the first
Born approximation as described in section \ref{sec2}, for impact energies 1-10 MeV \\
\end{tabular*}

\bigskip

\section*{Table 6.\label{tbl6te}
Ionization probabilities at zero impact parameter of Argon K, L and
M shells by 0.1-10 MeV protons. Explanation as in Table 5.}

\bigskip

\section*{Table 7.\label{tbl7te}
Ionization probabilities at zero impact parameter of Krypton M and N
shells by 0.1-10 MeV protons. Explanation as in Table 5.}

\bigskip

\section*{Table 8.\label{tbl8te}
Ionization probabilities at zero impact parameter of Xenon N and O
shells by 0.1-10 MeV protons. Explanation as in Table 5.}

\bigskip

\section*{Table 9.\label{tbl9te}
Mean radii $\left\langle r_{nl}\right\rangle $ of the atomic $nl$
subshells of Ne, Ar, Kr and Xe (in atomic units). }

\bigskip

\section*{Table 10.\label{tbl10te}
Impact parameter moment of order $-1$, $\left\langle
b_{nlm}^{-1}\right\rangle $ for 0.1-10 MeV protons in Ne, normalized
to the inverse of the mean radio of each subshell, $\left\langle
r_{nl}\right\rangle ^{-1}$. }
\begin{tabular*}{0.90\textwidth}{@{}@{\extracolsep{\fill}}lp{5.5in}@{}}
$E$ & proton impact energy  in MeV \\
$nlm$ & electron initial state \\
CDW-EIS & Impact parameter moment of order $-1$ given by Eq.
(\ref{150}), calculated within the
CDW-EIS approximation for impact energies 0.1-1 MeV \\
Born & Impact parameter moment of order $-1$ given
by Eq. (\ref{150}), calculated within the first Born approximation for impact energies 1-10 MeV \\
\end{tabular*}

\bigskip

\section*{Table 11.\label{tbl11te} Impact parameter moment of order $-1$, $\left\langle
b_{nlm}^{-1}\right\rangle $ for 0.1-10 MeV protons  in Ar,
normalized to the inverse of the mean radio of each subshell,
$\left\langle r_{nl}\right\rangle ^{-1}$. Explanation as in Table
10}

\bigskip

\section*{Table 12.\label{tbl12te}
Impact parameter moment of order $-1$, $\left\langle
b_{nlm}^{-1}\right\rangle $ for 0.1-10 MeV protons  in Kr,
normalized to the inverse of the mean radio of each subshell,
$\left\langle r_{nl}\right\rangle ^{-1}$. Explanation as in Table
10}

\bigskip

\section*{Table 13.\label{tbl13te}
Impact parameter moment of order $-1$, $\left\langle
b_{nlm}^{-1}\right\rangle $ for 0.1-10 MeV protons  in Xe,
normalized to the inverse of the mean radio of each subshell,
$\left\langle r_{nl}\right\rangle ^{-1}$. Explanation as in Table
10}

\bigskip

\section*{Table 14.\label{tbl14te}
Impact parameter moment of order $+1$, $\left\langle
b_{nlm}^{1}\right\rangle$, for proton in Ne, normalized to the mean
radio of each subshell, $\left\langle r_{nl}\right\rangle $. }
\begin{tabular*}{0.90\textwidth}{@{}@{\extracolsep{\fill}}lp{5.5in}@{}}
$E$ & proton impact energy  in MeV \\
$nlm$ & electron initial state \\
CDW-EIS & Impact parameter moment of order $+1$
given by Eq. (\ref{150}), calculated within the
CDW-EIS approximation for impact energies 0.1-1 MeV \\
Born & Impact parameter moment of order $+1$ given
by Eq. (\ref{150}), calculated within the first Born approximation for impact energies 1-10 MeV \\
\end{tabular*}

\bigskip

\section*{Table 15.\label{tbl15te}
Impact parameter moment of order $+1$, $\left\langle
b_{nlm}^{1}\right\rangle$, for proton in Ar, normalized to the mean
radio of each subshell, $\left\langle r_{nl}\right\rangle $.
Explanation as in Table 14. }

\bigskip

\section*{Table 16.\label{tbl16te}
Impact parameter moment of order $+1$, $\left\langle
b_{nlm}^{1}\right\rangle$, for proton in Kr, normalized to the mean
radio of each subshell, $\left\langle r_{nl}\right\rangle $.
Explanation as in Table 14. }

\bigskip

\section*{Table 17.\label{tbl17te}
Impact parameter moment of order $+1$, $\left\langle
b_{nlm}^{1}\right\rangle$, for proton in Xe, normalized to the mean
radio of each subshell, $\left\langle r_{nl}\right\rangle $.
Explanation as in Table 14. }

\newpage

\datatables 

\setlength{\LTleft}{0pt} \setlength{\LTright}{0pt}

\setlength{\tabcolsep}{0.62\tabcolsep}

\renewcommand{\arraystretch}{1.0}
\footnotesize

\begin{longtable}{@{\extracolsep\fill}lllllllllllllll@{}}
\caption{Total ionization cross sections, $\sigma_{nlm}$, of Neon K
and L shells by 0.1-10 MeV protons. Atomic units are used. See page
\pageref{tbl1te} for explanation.}\label{table1}
\mbox{Ne} & \mbox{CDW-EIS} &  &  &  &  &  &  & \ \ \ &\mbox{Born} &  &  &  &  &  \\
\hline \\
\endhead
E(MeV) & 0.1         & 0.2         & 0.3         & 0.4         & 0.5         & 0.7         & 1           & & 1           & 2           & 3           & 5           & 7           & 10          \\
\hline
$\sigma_{2p1}$   & 8.17$_{-1}$ & 7.96$_{-1}$ & 6.86$_{-1}$ & 5.95$_{-1}$ & 5.25$_{-1}$ & 4.26$_{-1}$ & 3.36$_{-1}$ & & 3.41$_{-1}$ & 2.04$_{-1}$ & 1.49$_{-1}$ & 9.93$_{-2}$ & 7.55$_{-2}$ & 5.63$_{-2}$ \\
$\sigma_{2p0}$    & 8.79$_{-1}$ & 9.03$_{-1}$ & 7.72$_{-1}$ & 6.59$_{-1}$ & 5.71$_{-1}$ & 4.52$_{-1}$ & 3.46$_{-1}$ & & 3.44$_{-1}$ & 1.99$_{-1}$ & 1.43$_{-1}$ & 9.35$_{-2}$ & 7.09$_{-2}$ & 5.19$_{-2}$ \\
$\sigma_{2s }$    & 2.49$_{-1}$ & 2.86$_{-1}$ & 2.40$_{-1}$ & 2.00$_{-1}$ & 1.69$_{-1}$ & 1.30$_{-1}$ & 9.62$_{-2}$ & & 9.46$_{-2}$ & 5.30$_{-2}$ & 3.74$_{-2}$ & 2.40$_{-2}$ & 1.78$_{-2}$ & 1.30$_{-2}$ \\
$\sigma_{1s }$    & 2.04$_{-5}$ & 1.44$_{-4}$ & 3.29$_{-4}$ & 5.20$_{-4}$ & 6.99$_{-4}$ & 9.92$_{-4}$ & 1.29$_{-3}$ & & 1.34$_{-3}$ & 1.58$_{-3}$ & 1.50$_{-3}$ & 1.25$_{-3}$ & 1.04$_{-3}$ & 8.35$_{-4}$ \\
\end{longtable}
\bigskip
\renewcommand{\arraystretch}{1.0}

\begin{longtable}{@{\extracolsep\fill}lllllllllllllll@{}}
\caption{Total ionization cross sections, $\sigma_{nlm}$, of  Argon
K, L and M shells by 0.1-10 MeV protons. Atomic units are used. See
page \pageref{tbl2te} for explanation}\label{table2}
\mbox{Ar} & \mbox{CDW-EIS} &  &  &  &  &  &  & \ \ \ &\mbox{Born} &  &  &  &  &  \\
\hline\\
\endhead
E(MeV)            & 0.1         & 0.2         & 0.3         & 0.4         & 0.5        & 0.6        & 1          & & 1           & 2           & 3           & 5           & 7           & 10          \\
\hline
$\sigma_{3p1}$    & 2.88        & 2.32        & 1.86        & 1.54        & 1.32       & 1.04       & 7.89$_{-1}$& & 8.02$_{-1}$ & 4.54$_{-1}$ & 3.21$_{-1}$ & 2.06$_{-1}$ & 1.53$_{-1}$ & 1.10$_{-1}$ \\
$\sigma_{3p0}$    & 3.55        & 2.70        & 2.06        & 1.66        & 1.39       & 1.06       & 7.83$_{-1}$& & 7.89$_{-1}$ & 4.32$_{-1}$ & 3.01$_{-1}$ & 1.90$_{-1}$ & 1.39$_{-1}$ & 9.96$_{-2}$ \\
$\sigma_{3s }$    & 6.69$_{-1}$ & 5.14$_{-1}$ & 3.81$_{-1}$ & 3.01$_{-1}$ &2.49$_{-1}$ &1.86$_{-1}$ & 1.36$_{-1}$& & 1.39$_{-1}$ & 7.34$_{-2}$ & 5.05$_{-2}$ & 3.14$_{-2}$ & 2.30$_{-2}$ & 1.65$_{-2}$ \\
$\sigma_{2p1}$    & 2.77$_{-3}$ & 7.88$_{-3}$ & 1.14$_{-2}$ & 1.35$_{-2}$ &1.48$_{-2}$ &1.57$_{-2}$ & 1.55$_{-2}$& & 1.64$_{-2}$ & 1.27$_{-2}$ & 1.02$_{-2}$ & 7.37$_{-3}$ & 5.84$_{-3}$ & 4.50$_{-3}$ \\
$\sigma_{2p0}$    & 2.16$_{-3}$ & 7.06$_{-3}$ & 1.14$_{-2}$ & 1.43$_{-2}$ &1.60$_{-2}$ &1.75$_{-2}$ & 1.76$_{-2}$& & 1.74$_{-2}$ & 1.38$_{-2}$ & 1.10$_{-2}$ & 7.81$_{-3}$ & 6.08$_{-3}$ & 4.60$_{-3}$ \\
$\sigma_{2s }$    & 1.17$_{-3}$ & 4.99$_{-3}$ & 7.63$_{-3}$ & 9.13$_{-3}$ &9.92$_{-3}$ &1.04$_{-2}$ & 9.94$_{-3}$& & 1.07$_{-2}$ & 7.58$_{-3}$ & 5.80$_{-3}$ & 3.99$_{-3}$ & 3.08$_{-3}$ & 2.31$_{-3}$ \\
$\sigma_{1s }$    & 2.10$_{-8}$ & 3.58$_{-7}$ & 1.45$_{-6}$ & 3.43$_{-6}$ &6.23$_{-6}$ &1.36$_{-5}$ & 2.68$_{-5}$& & 2.80$_{-5}$ & 6.98$_{-5}$ & 9.71$_{-5}$ & 1.21$_{-4}$ & 1.26$_{-4}$ & 1.20$_{-4}$ \\
\end{longtable}
\bigskip
\renewcommand{\arraystretch}{1.0}

\begin{longtable} {@{\extracolsep\fill}lllllllllllllll@{}}
\caption{Total ionization cross sections, $\sigma_{nlm}$, of
Krypton L, M and N shells by 0.1-10 MeV protons. Atomic units are
used. See page \pageref{tbl3te} for explanation.} \label{table3}
\mbox{Kr} & \mbox{CDW-EIS} &  &  &  &  &  &  & \ \ \ &\mbox{Born} &  &  &  &  &  \\
\hline\\
\endhead
E(MeV) & 0.1         & 0.2         & 0.3         & 0.4         & 0.5        & 0.7        & 1           & & 1 & 2 & 3 & 5 & 7 & 10 \\
\hline
$\sigma_{4p1}$    & 3.36$_{ }$  & 2.51$_{ }$  & 1.94$_{ }$  & 1.58$_{ }$  & 1.35$_{ }$ & 1.04$_{ }$ & 7.77$_{-1}$ &  & 8.01$_{-1}$ & 4.38$_{-1}$ & 3.09$_{-1}$ &1.98$_{-1}$& 1.47$_{-1}$ & 1.07$_{-1}$ \\
$\sigma_{4p0}$    & 4.48$_{ }$  & 2.91$_{ }$  & 2.12$_{ }$  & 1.76$_{ }$  & 1.38$_{ }$ & 1.03$_{ }$ & 7.53$_{-1}$ &  & 7.70$_{-1}$ & 4.24$_{-1}$ & 2.95$_{-1}$ &1.79$_{-1}$& 1.32$_{-1}$ & 9.52$_{-2}$ \\
$\sigma_{4s }$    & 9.92$_{-1}$ & 7.03$_{-1}$ & 5.09$_{-1}$ & 3.96$_{-1}$ &3.24$_{-1}$ &2.38$_{-1}$ & 1.70$_{-1}$ &  & 1.71$_{-1}$ & 8.80$_{-2}$ & 5.95$_{-2}$ & 3.63$_{-2}$ & 2.62$_{-2}$ & 1.88$_{-2}$ \\
$\sigma_{3d2}$    & 1.98$_{-2}$ & 3.53$_{-2}$ & 4.20$_{-2}$ & 4.45$_{-2}$ &4.50$_{-2}$ &4.37$_{-2}$ & 4.01$_{-2}$ &  & 4.25$_{-2}$ & 3.07$_{-2}$ & 2.43$_{-2}$ & 1.75$_{-2}$ & 1.39$_{-2}$ & 1.08$_{-2}$ \\
$\sigma_{3d1}$    & 2.42$_{-2}$ & 4.13$_{-2}$ & 4.85$_{-2}$ & 5.14$_{-2}$ &5.20$_{-2}$ &5.04$_{-2}$ & 4.59$_{-2}$ &  & 4.65$_{-2}$ & 3.31$_{-2}$ & 2.57$_{-2}$ & 1.80$_{-2}$ & 1.40$_{-2}$ & 1.06$_{-2}$ \\
$\sigma_{3d0}$    & 2.65$_{-2}$ & 4.81$_{-2}$ & 5.65$_{-2}$ & 5.88$_{-2}$ &5.86$_{-2}$ &5.54$_{-2}$ & 4.93$_{-2}$ &  & 4.79$_{-2}$ & 3.37$_{-2}$ & 2.61$_{-2}$ & 1.80$_{-2}$ & 1.39$_{-2}$ & 1.05$_{-2}$ \\
$\sigma_{3p1}$    & 3.36$_{-3}$ & 9.02$_{-3}$ & 1.24$_{-2}$ & 1.42$_{-2}$ &1.50$_{-2}$ &1.52$_{-2}$ & 1.41$_{-2}$ &  & 1.51$_{-2}$ & 1.04$_{-2}$ & 7.96$_{-3}$ & 5.48$_{-3}$ & 4.23$_{-3}$ & 3.19$_{-3}$ \\
$\sigma_{3p0}$    & 3.38$_{-3}$ & 9.92$_{-3}$ & 1.48$_{-2}$ & 1.70$_{-2}$ &1.77$_{-2}$ &1.74$_{-2}$ & 1.59$_{-2}$ &  & 1.63$_{-2}$ & 1.13$_{-2}$ & 8.46$_{-3}$ & 5.65$_{-3}$ & 4.28$_{-3}$ & 3.16$_{-3}$ \\
$\sigma_{3s }$    & 1.40$_{-3}$ & 5.63$_{-3}$ & 8.22$_{-3}$ & 9.44$_{-3}$ &1.00$_{-2}$ &1.02$_{-2}$ & 9.62$_{-3}$ &  & 1.03$_{-2}$ & 7.10$_{-3}$ & 5.38$_{-3}$ & 3.65$_{-3}$ & 2.79$_{-3}$ & 2.09$_{-3}$ \\
$\sigma_{2p1}$    & 5.88$_{-7}$ & 7.13$_{-6}$ & 2.28$_{-5}$ & 4.59$_{-5}$ &7.36$_{-5}$ &1.33$_{-4}$ & 2.14$_{-4}$ &  & 2.31$_{-4}$ & 3.71$_{-4}$ & 4.13$_{-4}$ & 4.13$_{-4}$ & 3.80$_{-4}$ & 3.31$_{-4}$ \\
$\sigma_{2p0}$    & 1.80$_{-6}$ & 1.16$_{-5}$ & 2.72$_{-5}$ & 4.59$_{-5}$ &6.54$_{-5}$ &1.03$_{-4}$ & 1.58$_{-4}$ &  & 1.80$_{-4}$ & 3.28$_{-4}$ & 4.00$_{-4}$ & 4.33$_{-4}$ & 4.12$_{-4}$ & 3.64$_{-4}$ \\
$\sigma_{2s }$    & 2.74$_{-7}$ & 1.40$_{-6}$ & 7.86$_{-6}$ & 2.21$_{-5}$ &4.30$_{-5}$ &9.70$_{-5}$ & 1.83$_{-4}$ &  & 1.90$_{-4}$ & 3.38$_{-4}$ & 3.73$_{-4}$ & 3.50$_{-4}$ & 3.09$_{-4}$ & 2.56$_{-4}$ \\

\end{longtable}

\bigskip
\renewcommand{\arraystretch}{1.0}
\footnotesize

\begin{longtable} {@{\extracolsep\fill}lllllllllllllll@{}}
\caption{Total ionization cross sections, $\sigma_{nlm}$, of Xenon
M, N and O shells by 0.1-10 MeV protons. Atomic units are used. See
page \pageref{tbl4te} for explanation.} \label{table4}
\mbox{Xe} & \mbox{CDW-EIS} &  &  &  &  &  &  & \ \ \ &\mbox{Born} &  &  &  &  &  \\
\hline\\
\endhead
E(MeV) & 0.1          & 0.2        & 0.3         & 0.4         & 0.5        & 0.7        & 1           &  & 1          & 2           & 3           & 5          & 7          & 10 \\
\hline
$\sigma_{5p1}$    & 5.09$_{ }$  & 3.52$_{ }$  & 2.67$_{ }$  & 2.16$_{ }$  &1.82$_{ } $ &1.40$_{  }$ & 1.05$_{ }$  &  & 1.06$_{ }$ & 5.91$_{-1}$ & 4.18$_{-1}$ &2.68$_{-1}$ &2.08$_{-1}$ & 1.46$_{-1}$ \\
$\sigma_{5p0}$    & 6.46$_{ }$  & 4.07$_{ }$  & 2.93$_{ }$  & 2.29$_{ }$  &1.89$_{ } $ &1.41$_{  }$ & 1.03$_{ }$  &  & 1.03$_{ }$ & 5.53$_{-1}$ & 3.85$_{-1}$ &2.42$_{-1}$ &1.80$_{-1}$ & 1.29$_{-1}$ \\
$\sigma_{5s }$    & 1.72$_{ }$  & 1.05$_{ }$  & 7.37$_{-1}$ & 5.67$_{-1}$ &4.61$_{-1}$ &3.36$_{-1}$ & 2.39$_{-1}$ &  & 2.39$_{-1}$ & 1.22$_{-1}$ & 8.23$_{-2}$ &5.00$_{-2}$& 3.60$_{-2}$ & 2.54$_{-2}$ \\
$\sigma_{4d2}$    & 1.80$_{-1}$ & 2.35$_{-1}$ & 2.33$_{-1}$ & 2.20$_{-1}$ &2.06$_{-1}$ &1.79$_{-1}$ & 1.50$_{-1}$ &  & 1.55$_{-1}$ & 9.96$_{-2}$ & 7.52$_{-2}$ & 5.17$_{-2}$ & 3.99$_{-2}$ & 3.03$_{-2}$ \\
$\sigma_{4d1}$    & 2.12$_{-1}$ & 2.77$_{-1}$ & 2.72$_{-1}$ & 2.54$_{-1}$ &2.35$_{-1}$ &2.00$_{-1}$ & 1.63$_{-1}$ &  & 1.67$_{-1}$ & 1.03$_{-1}$ & 7.59$_{-2}$ & 5.08$_{-2}$ & 3.87$_{-2}$ & 2.90$_{-2}$ \\
$\sigma_{4d0}$    & 2.16$_{-1}$ & 3.01$_{-1}$ & 2.99$_{-1}$ & 2.77$_{-1}$ &2.53$_{-1}$ &2.12$_{-1}$ & 1.70$_{-1}$ &  & 1.72$_{-1}$ & 1.04$_{-1}$ & 7.59$_{-2}$ & 5.03$_{-2}$ & 3.81$_{-2}$ & 2.83$_{-2}$ \\
$\sigma_{4p1}$    & 1.67$_{-2}$ & 2.77$_{-2}$ & 2.99$_{-2}$ & 2.96$_{-2}$ &2.84$_{-2}$ &2.54$_{-2}$ & 2.12$_{-2}$ &  & 2.21$_{-2}$ & 1.37$_{-2}$ & 1.00$_{-2}$ & 6.65$_{-3}$ & 5.02$_{-3}$ & 3.71$_{-3}$ \\
$\sigma_{4p0}$    & 1.33$_{-2}$ & 3.67$_{-2}$ & 4.21$_{-2}$ & 3.99$_{-2}$ &3.65$_{-2}$ &3.05$_{-2}$ & 2.46$_{-2}$ &  & 2.49$_{-2}$ & 1.45$_{-2}$ & 1.02$_{-2}$ & 6.44$_{-3}$ & 4.73$_{-3}$ & 3.39$_{-3}$ \\
$\sigma_{4s }$    & 9.35$_{-3}$ & 1.92$_{-2}$ & 1.98$_{-2}$ & 1.93$_{-2}$ &1.86$_{-2}$ &1.71$_{-2}$ & 1.48$_{-2}$ &  & 1.51$_{-2}$ & 9.48$_{-3}$ & 6.96$_{-3}$ & 4.60$_{-3}$ & 3.47$_{-3}$ & 2.56$_{-3}$ \\
$\sigma_{3d2}$    & 6.01$_{-5}$ & 3.14$_{-4}$ & 6.33$_{-4}$ & 9.33$_{-4}$ &1.19$_{-3}$ &1.57$_{-3}$ & 1.89$_{-3}$ &  & 2.00$_{-3}$ & 2.20$_{-3}$ & 2.09$_{-3}$ & 1.78$_{-3}$ & 1.54$_{-3}$ & 1.28$_{-3}$ \\
$\sigma_{3d1}$    & 7.22$_{-5}$ & 2.59$_{-4}$ & 4.64$_{-4}$ & 6.62$_{-4}$ &8.53$_{-4}$ &1.22$_{-3}$ & 1.68$_{-3}$ &  & 1.83$_{-3}$ & 2.31$_{-3}$ & 2.30$_{-3}$ & 2.00$_{-3}$ & 1.71$_{-3}$ & 1.40$_{-3}$ \\
$\sigma_{3d0}$    & 7.46$_{-5}$ & 2.62$_{-4}$ & 4.32$_{-4}$ & 5.94$_{-4}$ &7.67$_{-4}$ &1.12$_{-3}$ & 1.56$_{-3}$ &  & 1.73$_{-3}$ & 2.34$_{-3}$ & 2.36$_{-3}$ & 2.08$_{-3}$ & 1.78$_{-3}$ & 1.44$_{-3}$ \\
$\sigma_{3p1}$    & 6.23$_{-6}$ & 7.72$_{-5}$ & 2.37$_{-4}$ & 4.40$_{-4}$ &6.48$_{-4}$ &9.99$_{-4}$ & 1.31$_{-3}$ &  & 1.27$_{-3}$ & 1.42$_{-3}$ & 1.30$_{-3}$ & 1.04$_{-3}$ & 8.53$_{-4}$ & 6.76$_{-4}$ \\
$\sigma_{3p0}$    & 1.73$_{-5}$ & 9.49$_{-5}$ & 2.02$_{-4}$ & 3.31$_{-4}$ &4.63$_{-4}$ &7.04$_{-4}$ & 1.03$_{-3}$ &  & 1.08$_{-3}$ & 1.42$_{-3}$ & 1.38$_{-3}$ & 1.14$_{-3}$ & 9.39$_{-4}$ & 7.36$_{-4}$ \\
$\sigma_{3s }$    & 3.37$_{-6}$ & 2.55$_{-5}$ & 1.13$_{-4}$ & 2.53$_{-4}$ &4.17$_{-4}$ &7.38$_{-4}$ & 1.06$_{-3}$ &  & 9.86$_{-4}$ & 1.14$_{-3}$ & 1.04$_{-3}$ & 8.14$_{-4}$ & 6.60$_{-4}$ & 5.15$_{-4}$ \\
\end{longtable}


\renewcommand{\arraystretch}{1.0}
\footnotesize

\begin{longtable}{@{\extracolsep\fill}lllllllllllllll@{}}
\caption{Ionization probabilities at zero impact parameter of Neon K
and L shells by 0.1-10 MeV protons. See page \pageref{tbl6te} for
explanation.} \label{table6}
\mbox{Ne}& \mbox{CDW-EIS}&          &              &             &             &             &      & \ \ \ &\mbox{Born}   &             &            &              &            &               \\
\hline\\
\endhead
E(MeV) & 0.1       & 0.2        & 0.3          & 0.4         & 0.5         &0.7          & 1.0         && 1            & 2           & 3           & 5           & 7           & 10   \\
\hline
$P _{2p1}(b=0)$  &\ 8.50$_{-2}$ & 8.17$_{-2}$ & 6.97$_{-2}$ & 5.98$_{-2}$ & 5.20$_{-2}$ & 4.11$_{-2}$ & 3.10$_{-2}$ && 3.40$_{-2}$  & 1.75$_{-2}$ & 1.18$_{-2}$ & 7.06$_{-3}$ & 5.03$_{-3}$ & 3.51$_{-3}$  \\
$P _{2p0}(b=0)$ &\ 2.19$_{-1}$ & 1.92$_{-1}$ & 1.51$_{-1}$ & 1.23$_{-1}$ & 1.02$_{-1}$ & 7.60$_{-2}$ & 5.38$_{-2}$ && 5.80$_{-2}$  & 2.73$_{-2}$ & 1.77$_{-2}$ & 1.02$_{-2}$ & 7.17$_{-3}$ & 4.94$_{-3}$  \\
$P _{2s }\ (b=0)$ &\ 3.97$_{-2}$ & 4.47$_{-2}$ & 4.27$_{-2}$ & 3.97$_{-2}$ & 3.67$_{-2}$ & 3.18$_{-2}$ & 2.62$_{-2}$ && 2.63$_{-2}$  & 1.60$_{-2}$ & 1.14$_{-2}$ & 7.11$_{-3}$ & 5.15$_{-3}$ & 3.63$_{-3}$  \\
$P _{1s }\ (b=0)$ &\ 6.78$_{-4}$ & 3.15$_{-3}$ & 5.81$_{-3}$ & 7.87$_{-3}$ & 9.47$_{-3}$ & 1.13$_{-2}$ & 1.22$_{-2}$ && 1.28$_{-2}$  & 1.18$_{-2}$ & 1.01$_{-2}$ & 7.32$_{-3}$ & 5.60$_{-3}$ & 4.09$_{-3}$  \\
\end{longtable}


\renewcommand{\arraystretch}{1.0}
\footnotesize

\begin{longtable}{@{\extracolsep\fill}lllllllllllllll@{}}
\caption{Ionization probabilities at zero impact parameter of Argon
K, L and M shells by 0.1-10 MeV protons. See page \pageref{tbl7te}
for explanation.} \label{table7}
\mbox{Ar}  & \mbox{CDW-EIS}  & & & & & & & \ \ \ &\mbox{Born}  & & & & &    \\
\hline\\
\endhead
E(MeV) & 0.1      & 0.2         & 0.3         & 0.4         & 0.5         & 0.7         & 1            &   & 1            & 2            & 3           & 5          & 7              & 10        \\
\hline
$P _{3p1}(b=0)$  & 9.51$_{-2}$ & 6.78$_{-2}$ & 5.21$_{-2}$ & 4.23$_{-2}$ & 3.56$_{-2}$ & 2.72$_{-2}$ & 2.02$_{-2}$  &   & 2.09$_{-2}$   & 1.12$_{-2}$  & 7.62$_{-3}$ & 4.66$_{-3}$ &   3.36$_{-3}$  & 2.36$_{-3}$ \\
$P _{3p0}(b=0)$  & 2.85$_{-1}$ & 1.77$_{-1}$ & 1.28$_{-1}$ & 1.01$_{-1}$ & 8.38$_{-2}$ & 6.25$_{-2}$ & 4.52$_{-2}$  &   & 4.53$_{-2}$   & 2.32$_{-2}$  & 1.54$_{-2}$ & 9.16$_{-3}$ &   6.49$_{-3}$  & 4.55$_{-3}$ \\
$P _{3s }\ (b=0)$& 5.67$_{-2}$ & 5.92$_{-2}$ & 4.99$_{-2}$ & 4.11$_{-2}$ & 3.44$_{-2}$ & 2.57$_{-2}$ & 1.87$_{-2}$  &   & 1.93$_{-2}$   & 1.03$_{-2}$  & 7.08$_{-3}$ & 4.41$_{-3}$ &   3.21$_{-3}$  & 2.27$_{-3}$ \\
$P _{2p1}(b=0)$  & 4.02$_{-3}$ & 1.09$_{-2}$ & 1.47$_{-2}$ & 1.64$_{-2}$ & 1.71$_{-2}$ & 1.71$_{-2}$ & 1.57$_{-2}$  &   & 1.60$_{-2}$   & 1.08$_{-2}$  & 7.96$_{-3}$ & 5.17$_{-3}$ &   3.81$_{-3}$  & 2.70$_{-3}$ \\
$P _{2p0}(b=0)$  & 8.58$_{-3}$ & 3.02$_{-2}$ & 4.51$_{-2}$ & 5.36$_{-2}$ & 5.72$_{-2}$ & 5.73$_{-2}$ & 5.12$_{-2}$  &   & 4.56$_{-2}$   & 2.74$_{-2}$  & 1.87$_{-2}$ & 1.10$_{-2}$ &   7.65$_{-3}$  & 5.19$_{-3}$ \\
$P _{2s }\ (b=0)$& 3.83$_{-3}$ & 1.08$_{-2}$ & 1.26$_{-2}$ & 1.22$_{-2}$ & 1.12$_{-2}$ & 9.34$_{-3}$ & 7.96$_{-3}$  &   & 8.42$_{-3}$   & 7.27$_{-3}$  & 6.46$_{-3}$ & 5.00$_{-3}$ &   3.98$_{-3}$  & 3.01$_{-3}$ \\
$P _{1s }\ (b=0)$& 8.15$_{-6}$ & 7.81$_{-5}$ & 2.21$_{-4}$ & 4.09$_{-4}$ & 6.29$_{-4}$ & 1.12$_{-3}$ & 1.80$_{-3}$  &   & 1.91$_{-3}$   & 3.23$_{-3}$  & 3.68$_{-3}$ & 3.76$_{-3}$ &   3.58$_{-3}$  & 3.11$_{-3}$ \\
\end{longtable}

\renewcommand{\arraystretch}{1.0}
\footnotesize

\begin{longtable}{@{\extracolsep\fill}lllllllllllllll@{}}
\caption{Ionization probabilities at zero impact parameter of
Krypton M and N shells by 0.1-10 MeV protons.  See page
\pageref{tbl8te} for explanation.} \label{table8}
\mbox{Kr}  & \mbox{CDW-EIS} & & & & & & & \ \ \ &\mbox{Born} & & & & & \\
\hline
\\
\endhead
E(MeV) & 0.1      & 0.2         & 0.3         & 0.4         & 0.5         & 0.7         & 1              &  & 1            & 2            & 3           & 5          & 7              & 10          \\
\hline
$P _{4p1}(b=0)$   &\ 1.04$_{-1}$  & 6.64$_{-2}$ & 4.88$_{-2}$ & 3.86$_{-2}$ & 3.19$_{-2}$ & 2.38$_{-2}$ & 1.72$_{-2}$  &  & 1.74$_{-2}$ & 9.08$_{-3}$ & 6.17$_{-3}$ & 3.77$_{-3}$ &  2.72$_{-3}$ & 1.92$_{-3}$  \\
$P _{4p0}(b=0)$   &\ 3.57$_{-1}$  & 1.89$_{-1}$ & 1.32$_{-1}$ & 1.02$_{-1}$ & 8.22$_{-2}$ & 5.91$_{-2}$ & 4.10$_{-2}$  &  & 4.11$_{-2}$ & 2.03$_{-2}$ & 1.36$_{-2}$ & 8.19$_{-3}$ &  5.86$_{-3}$ & 4.12$_{-3}$  \\
$P _{4s }\ (b=0)$ &\ 5.62$_{-2}$  & 6.31$_{-2}$ & 4.99$_{-2}$ & 3.98$_{-2}$ & 3.29$_{-2}$ & 2.48$_{-2}$ & 1.84$_{-2}$  &  & 1.84$_{-2}$ & 1.01$_{-2}$ & 6.90$_{-3}$ & 4.21$_{-3}$ &  3.04$_{-3}$ & 2.15$_{-3}$  \\
$P _{3d2}(b=0)$   &\ 1.31$_{-2}$  & 2.06$_{-2}$ & 2.29$_{-2}$ & 2.32$_{-2}$ & 2.24$_{-2}$ & 2.03$_{-2}$ & 1.73$_{-2}$  &  & 1.79$_{-2}$ & 1.11$_{-2}$ & 7.99$_{-3}$ & 5.11$_{-3}$ &  3.75$_{-3}$ & 2.67$_{-3}$  \\
$P _{3d1}(b=0)$   &\ 2.50$_{-2}$  & 4.16$_{-2}$ & 4.66$_{-2}$ & 4.67$_{-2}$ & 4.48$_{-2}$ & 3.97$_{-2}$ & 3.25$_{-2}$  &  & 3.13$_{-2}$ & 1.82$_{-2}$ & 1.25$_{-2}$ & 7.45$_{-3}$ &  5.25$_{-3}$ & 3.61$_{-3}$  \\
$P _{3d0}(b=0)$   &\ 4.67$_{-2}$  & 8.28$_{-2}$ & 9.40$_{-2}$ & 9.46$_{-2}$ & 9.07$_{-2}$ & 7.88$_{-2}$ & 6.24$_{-2}$  &  & 5.57$_{-2}$ & 2.94$_{-2}$ & 1.93$_{-2}$ & 1.10$_{-2}$ &  7.60$_{-3}$ & 5.15$_{-3}$  \\
$P _{3p1}(b=0)$   &\ 2.94$_{-3}$  & 5.69$_{-3}$ & 6.65$_{-3}$ & 6.77$_{-3}$ & 6.58$_{-3}$ & 6.01$_{-3}$ & 5.33$_{-3}$  &  & 5.49$_{-3}$ & 4.40$_{-3}$ & 3.70$_{-3}$ & 2.82$_{-3}$ &  2.27$_{-3}$ & 1.75$_{-3}$  \\
$P _{3p0}(b=0)$   &\ 9.69$_{-3}$  & 2.44$_{-2}$ & 3.15$_{-2}$ & 3.24$_{-2}$ & 3.04$_{-2}$ & 2.50$_{-2}$ & 1.88$_{-2}$  &  & 1.64$_{-2}$ & 1.22$_{-2}$ & 1.05$_{-2}$ & 7.89$_{-3}$ &  6.13$_{-3}$ & 4.51$_{-3}$  \\
$P _{3s }\ (b=0)$ &\ 2.53$_{-3}$  & 6.50$_{-3}$ & 7.52$_{-3}$ & 7.30$_{-3}$ & 7.16$_{-3}$ & 7.46$_{-3}$ & 8.23$_{-3}$  &  & 8.67$_{-3}$ & 6.32$_{-3}$ & 4.55$_{-3}$ & 2.86$_{-3}$ &  2.13$_{-3}$ & 1.59$_{-3}$  \\
\end{longtable}


\renewcommand{\arraystretch}{1.0}
\footnotesize

\begin{longtable}{@{\extracolsep\fill}lllllllllllllll@{}}
\caption{Ionization probabilities at zero impact parameter of Xenon
N and O shells by 0.1-10 MeV protons.  See page \pageref{tbl9te} for
explanation.} \label{table9}
\mbox{Xe}  & \mbox{CDW-EIS} & & & & & &  & \ \ \ &\mbox{Born} & & & & &    \\
\hline
\\
\endhead
E(MeV)     & 0.1   & 0.2           & 0.3       & 0.4          & 0.5          & 0.7          & 1           & &1           & 2           & 3           & 5           &  7 & 10  \\
\hline
$P _{5p1}(b=0)$    & 9.85$_{-2}$ & 6.71$_{-2}$  & 4.91$_{-2}$ & 3.85$_{-2}$  & 3.17$_{-2}$  & 2.34$_{-2}$  & 1.69$_{-2}$ & &1.70$_{-2}$ & 8.86$_{-3}$ & 6.02$_{-3}$ & 3.68$_{-3}$ & 2.66$_{-3}$ & 1.88$_{-3}$  \\
$P _{5p0}(b=0)$    & 3.56$_{-1}$ & 2.19$_{-1}$  & 1.48$_{-1}$ & 1.08$_{-1}$  & 8.49$_{-2}$  & 5.97$_{-2}$  & 4.19$_{-2}$ & &4.23$_{-2}$ & 2.12$_{-2}$ & 1.41$_{-2}$ & 8.65$_{-3}$ & 6.06$_{-3}$ & 4.16$_{-3}$  \\
$P _{5s }\ (b=0)$  & 9.14$_{-2}$ & 5.46$_{-2}$  & 4.16$_{-2}$ & 3.54$_{-2}$  & 3.12$_{-2}$  & 2.48$_{-2}$  & 1.87$_{-2}$ & &1.86$_{-2}$ & 9.89$_{-3}$ & 6.81$_{-3}$ & 4.20$_{-3}$ & 3.02$_{-3}$ & 2.12$_{-3}$  \\
$P _{4d2}(b=0)$    & 3.70$_{-2}$ & 3.48$_{-2}$  & 2.95$_{-2}$ & 2.53$_{-2}$  & 2.21$_{-2}$  & 1.77$_{-2}$  & 1.36$_{-2}$ & &1.39$_{-2}$ & 8.14$_{-3}$ & 5.83$_{-3}$ & 3.75$_{-3}$ & 2.78$_{-3}$ & 2.00$_{-3}$  \\
$P _{4d1}(b=0)$    & 1.09$_{-1}$ & 9.76$_{-2}$  & 7.55$_{-2}$ & 6.00$_{-2}$  & 4.90$_{-2}$  & 3.53$_{-2}$  & 2.48$_{-2}$ & &2.40$_{-2}$ & 1.31$_{-2}$ & 9.26$_{-3}$ & 5.84$_{-3}$ & 4.24$_{-3}$ & 2.99$_{-3}$  \\
$P _{4d0}(b=0)$    & 2.02$_{-1}$ & 1.96$_{-1}$  & 1.59$_{-1}$ & 1.24$_{-1}$  & 9.88$_{-2}$  & 6.83$_{-2}$  & 4.54$_{-2}$ & &4.03$_{-2}$ & 2.32$_{-2}$ & 1.73$_{-2}$ & 1.11$_{-2}$ & 8.00$_{-3}$ & 5.47$_{-3}$  \\
$P _{4p1}(b=0)$    & 5.08$_{-3}$ & 7.47$_{-3}$  & 7.32$_{-3}$ & 6.88$_{-3}$  & 6.53$_{-3}$  & 6.02$_{-3}$  & 5.46$_{-3}$ & &5.64$_{-3}$ & 4.03$_{-3}$ & 3.11$_{-3}$ & 2.16$_{-3}$ & 1.68$_{-3}$ & 1.27$_{-3}$  \\
$P _{4p0}(b=0)$    & 2.32$_{-2}$ & 3.96$_{-2}$  & 3.65$_{-2}$ & 3.10$_{-2}$  & 2.69$_{-2}$  & 2.25$_{-2}$  & 2.11$_{-2}$ & &2.10$_{-2}$ & 1.30$_{-2}$ & 8.74$_{-3}$ & 5.30$_{-3}$ & 3.91$_{-3}$ & 2.85$_{-3}$  \\
$P _{4s }\ (b=0)$  & 5.46$_{-3}$ & 8.71$_{-3}$  & 9.12$_{-3}$ & 9.87$_{-3}$  & 1.04$_{-2}$  & 1.07$_{-2}$  & 8.99$_{-3}$ & &7.74$_{-3}$ & 4.79$_{-3}$ & 3.79$_{-3}$ & 2.66$_{-3}$ & 2.01$_{-3}$ & 1.45$_{-3}$  \\
\end{longtable}



\renewcommand{\arraystretch}{1.0}

\begin{longtable}{ccccccc}
\caption{Mean radii $\left\langle r_{nl}\right\rangle $ of the
atomic $nl$ subshells of Ne, Ar, Kr and Xe (in atomic
units).}\label{table5}
\mbox{ Ne}        & \ \ \ \ \  & \mbox{Ar}      &\ \ \  \ \ & \mbox{Kr}      &\  \ \ \ \ &  \mbox{Xe}        \\
\hline\\
\endhead
\ $\left\langle r_{2p}\right\rangle $ = 0.9653  &  & $\left\langle r_{3p}\right\rangle $ = 1.6629  &  & $\left\langle r_{4p}\right\rangle $ = 1.9516 &  & $\left\langle r_{5p}\right\rangle $ = 2.3380  \\
\ $\left\langle r_{2s}\right\rangle $ = 0.8921  &  & $\left\langle r_{3s}\right\rangle $ = 1.4222  &  & $\left\langle r_{4s}\right\rangle $ = 1.6294 &  & $\left\langle r_{5s}\right\rangle $ = 1.9810  \\
\ $\left\langle r_{1s}\right\rangle $ = 0.1576  &  & $\left\langle r_{2p}\right\rangle $ = 0.3753  &  & $\left\langle r_{3d}\right\rangle $ = 0.5509 &  & $\left\langle r_{4d}\right\rangle $ = 0.8704  \\
                   &  & $\left\langle r_{2s}\right\rangle $ = 0.4123  &  & $\left\langle r_{3p}\right\rangle $ = 0.5426 &  & $\left\langle r_{4p}\right\rangle $ = 0.7770  \\
                   &  & $\left\langle r_{1s}\right\rangle $ = 0.0861  &  & $\left\langle r_{3s}\right\rangle $ = 0.5378 &  & $\left\langle r_{4s}\right\rangle $ = 0.7453  \\
\end{longtable}
\newpage

%
%

\renewcommand{\arraystretch}{1.0}

\footnotesize

\begin{longtable}{@{\extracolsep\fill}lllllllllllllll@{}}
\caption{Impact parameter moment of order $-1$, $\left\langle
b_{nlm}^{-1}\right\rangle $ for proton impact in Ne, normalized to
the inverse of the mean radio of each subshell, $\left\langle
r_{nl}\right\rangle ^{-1}$.  See page \pageref{tbl10te} for
explanation.} \label{table10}

\mbox{ Ne} & \mbox{CDW-EIS} & & & & & &  & \ \ \ &\mbox{Born} & & & & &    \\
\hline
\\
\endhead
E(MeV)  & 0.1     & 0.2         & 0.3         & 0.4         & 0.5         & 0.7         & 1           & & 1           & 2           & 3           & 5           &  7 & 10   \\
\hline
2p1 & 9.15$_{-1}$ & 9.17$_{-1}$ & 9.08$_{-1}$ & 8.95$_{-1}$ & 8.82$_{-1}$ & 8.57$_{-1}$ & 8.26$_{-1}$ & & 8.36$_{-1}$ & 7.58$_{-1}$ & 7.16$_{-1}$ & 6.65$_{-1}$ & 6.35$_{-1}$   & 6.05$_{-1}$ \\
2p0 & 1.34$_{  }$ & 1.23$_{  }$ & 1.18$_{  }$ & 1.15$_{  }$ & 1.11$_{  }$ & 1.07$_{  }$ & 1.01$_{  }$ & & 1.02$_{  }$ & 8.97$_{-1}$ & 8.37$_{-1}$ & 7.68$_{-1}$ & 7.29$_{-1}$   & 6.91$_{-1}$ \\
2s  & 1.19$_{  }$ & 1.17$_{  }$ & 1.20$_{  }$ & 1.23$_{  }$ & 1.25$_{  }$ & 1.29$_{  }$ & 1.31$_{  }$ & & 1.32$_{  }$ & 1.31$_{  }$ & 1.29$_{  }$ & 1.25$_{  }$ & 1.23$_{  }$   & 1.19$_{  }$ \\
1s  & 2.78$_{  }$ & 2.26$_{  }$ & 2.02$_{  }$ & 1.86$_{  }$ & 1.76$_{  }$ & 1.60$_{  }$ & 1.45$_{  }$ & & 1.46$_{  }$ & 1.28$_{  }$ & 1.20$_{  }$ & 1.12$_{  }$ & 1.06$_{  }$   & 1.01$_{  }$ \\
\end{longtable}
\bigskip

\renewcommand{\arraystretch}{1.0}
\footnotesize

\begin{longtable}{@{\extracolsep\fill}lllllllllllllll@{}}
\caption{Impact parameter moment of order $-1$, $\left\langle
b_{nlm}^{-1}\right\rangle$, for proton impact in Ar, normalized to
the inverse of the mean radio of each subshell, $\left\langle
r_{nl}\right\rangle ^{-1}$. See page \pageref{tbl11te} for
explanation.} \label{table11}

\mbox{ Ar}  & \mbox{CDW-EIS} & & & & & &  & \ \ \ &\mbox{Born} & & & & &    \\
\hline
\\
\endhead
E(MeV) & 0.1      & 0.2         & 0.3         & 0.4         & 0.5         & 0.7         & 1            & & 1           & 2           & 3           & 5           &  7 & 10  \\
\hline
3p1 & 1.02$_{  }$ & 9.76$_{-1}$ & 9.48$_{-1}$ & 9.25$_{-1}$ & 9.05$_{-1}$ & 8.76$_{-1}$ & 8.45$_{-1}$  & &  8.53$_{-1}$    & 7.90$_{-1}$ & 7.60$_{-1}$ & 7.22$_{-1}$ &   7.03$_{-1}$  & 6.85$_{-1}$ \\
3p0 & 1.39$_{  }$ & 1.29$_{  }$ & 1.24$_{  }$ & 1.20$_{  }$ & 1.17$_{  }$ & 1.13$_{  }$ & 1.08$_{  }$  & &  1.08$_{  }$    & 9.93$_{-1}$ & 9.50$_{-1}$ & 9.00$_{-1}$ &   8.75$_{-1}$  & 8.56$_{-1}$ \\
3s  & 1.37$_{  }$ & 1.48$_{  }$ & 1.55$_{  }$ & 1.57$_{  }$ & 1.58$_{  }$ & 1.59$_{  }$ & 1.59$_{  }$  & &  1.60$_{  }$    & 1.58$_{  }$ & 1.56$_{  }$ & 1.53$_{  }$ &   1.51$_{  }$  & 1.48$_{  }$ \\
2p1 & 1.60$_{  }$ & 1.45$_{  }$ & 1.37$_{  }$ & 1.31$_{  }$ & 1.26$_{  }$ & 1.21$_{  }$ & 1.16$_{  }$  & &  1.16$_{  }$    & 1.07$_{  }$ & 1.02$_{  }$ & 9.60$_{-1}$ &   9.18$_{-1}$  & 8.73$_{-1}$ \\
2p0 & 2.27$_{  }$ & 2.32$_{  }$ & 2.21$_{  }$ & 2.13$_{  }$ & 2.06$_{  }$ & 1.94$_{  }$ & 1.80$_{  }$  & &  1.73$_{  }$    & 1.48$_{  }$ & 1.36$_{  }$ & 1.23$_{  }$ &   1.15$_{  }$  & 1.08$_{  }$ \\
2s  & 2.29$_{  }$ & 1.90$_{  }$ & 1.68$_{  }$ & 1.53$_{  }$ & 1.43$_{  }$ & 1.31$_{  }$ & 1.25$_{  }$  & &  1.26$_{  }$    & 1.27$_{  }$ & 1.29$_{  }$ & 1.30$_{  }$ &   1.29$_{  }$  & 1.27$_{  }$ \\
1s  & 3.38$_{  }$ & 3.17$_{  }$ & 2.81$_{  }$ & 2.58$_{  }$ & 2.44$_{  }$ & 2.22$_{  }$ & 2.02$_{  }$  & &  2.03$_{  }$    & 1.68$_{  }$ & 1.53$_{  }$ & 1.38$_{  }$ &   1.31$_{  }$  & 1.23$_{  }$ \\
\end{longtable}
\bigskip

\renewcommand{\arraystretch}{1.0}

\footnotesize

\begin{longtable}{@{\extracolsep\fill}lllllllllllllll@{}}
\caption{Impact parameter moment of order $-1$, $\left\langle
b_{nlm}^{-1}\right\rangle$, for proton impact in  Kr, normalized to
the inverse of the  mean radio of each subshell, $\left\langle
r_{nl}\right\rangle ^{-1}$. See page \pageref{tbl12te} for
explanation.} \label{table12}

\mbox{Kr}  & \mbox{CDW-EIS} & & & & & &  & \ \ \ &\mbox{Born} & & & & &    \\
\hline
\\
\endhead
E(MeV) & 0.1        & 0.2         & 0.3         & 0.4         & 0.5         & 0.7         & 1           &  &1           & 2           & 3           & 5           &  7 & 10  \\
\hline
4p1 &\ 1.15$_{  }$  & 1.09$_{  }$ & 1.06$_{  }$ & 1.03$_{  }$ & 1.01$_{  }$ & 9.83$_{-1}$ & 9.53$_{-1}$ &  &9.47$_{-1}$ & 8.88$_{-1}$ & 8.55$_{-1}$ & 8.15$_{-1}$ &  7.91$_{-1}$ & 7.68$_{-1}$ \\
4p0 &\ 1.62$_{  }$  & 1.48$_{  }$ & 1.42$_{  }$ & 1.38$_{  }$ & 1.35$_{  }$ & 1.31$_{  }$ & 1.26$_{  }$ &  &1.25$_{  }$ & 1.16$_{  }$ & 1.12$_{  }$ & 1.06$_{  }$ &  1.03$_{  }$ & 9.99$_{-1}$ \\
4s  &\ 1.35$_{  }$  & 1.50$_{  }$ & 1.55$_{  }$ & 1.58$_{  }$ & 1.59$_{  }$ & 1.61$_{  }$ & 1.63$_{  }$ &  &1.62$_{  }$ & 1.63$_{  }$ & 1.63$_{  }$ & 1.62$_{  }$ &  1.62$_{  }$ & 1.61$_{  }$ \\
3d2 &\ 1.46$_{  }$  & 1.31$_{  }$ & 1.24$_{  }$ & 1.20$_{  }$ & 1.17$_{  }$ & 1.12$_{  }$ & 1.08$_{  }$ &  &1.07$_{  }$ & 9.92$_{-1}$ & 9.43$_{-1}$ & 8.81$_{-1}$ &  8.40$_{-1}$ & 7.99$_{-1}$ \\
3d1 &\ 1.77$_{  }$  & 1.69$_{  }$ & 1.62$_{  }$ & 1.55$_{  }$ & 1.49$_{  }$ & 1.41$_{  }$ & 1.33$_{  }$ &  &1.30$_{  }$ & 1.17$_{  }$ & 1.09$_{  }$ & 9.99$_{-1}$ &  9.44$_{-1}$ & 8.89$_{-1}$ \\
3d0 &\ 2.14$_{  }$  & 2.10$_{  }$ & 2.04$_{  }$ & 1.97$_{  }$ & 1.90$_{  }$ & 1.79$_{  }$ & 1.65$_{  }$ &  &1.57$_{  }$ & 1.34$_{  }$ & 1.23$_{  }$ & 1.11$_{  }$ &  1.04$_{  }$ & 9.75$_{-1}$ \\
3p1 &\ 1.74$_{  }$  & 1.51$_{  }$ & 1.38$_{  }$ & 1.29$_{  }$ & 1.24$_{  }$ & 1.18$_{  }$ & 1.15$_{  }$ &  &1.15$_{  }$ & 1.15$_{  }$ & 1.15$_{  }$ & 1.16$_{  }$ &  1.15$_{  }$ & 1.14$_{  }$ \\
3p0 &\ 2.72$_{  }$  & 2.50$_{  }$ & 2.34$_{  }$ & 2.22$_{  }$ & 2.10$_{  }$ & 1.93$_{  }$ & 1.76$_{  }$ &  &1.69$_{  }$ & 1.64$_{  }$ & 1.65$_{  }$ & 1.66$_{  }$ &  1.64$_{  }$ & 1.61$_{  }$ \\
3s  &\ 2.29$_{  }$  & 1.90$_{  }$ & 1.70$_{  }$ & 1.56$_{  }$ &
1.49$_{  }$ & 1.44$_{  }$ & 1.46$_{  }$ &  &1.44$_{  }$ & 1.45$_{ }$
& 1.44$_{  }$ & 1.41$_{  }$ &  1.39$_{  }$ & 1.36$_{  }$ \\
\end{longtable}
\bigskip

\renewcommand{\arraystretch}{1.0}

\begin{longtable}{@{\extracolsep\fill}lllllllllllllll@{}}
\caption{Impact parameter moment of order $-1$, $\left\langle
b_{nlm}^{-1}\right\rangle$ , for proton impact in  Xe, normalized to
the inverse of the mean radio of each subshell, $\left\langle
r_{nl}\right\rangle ^{-1}$. See page \pageref{tbl13te} for
explanation.} \label{table13}

\mbox{ Xe}  & \mbox{CDW-EIS} & & & & & &  & \ \ \ &\mbox{Born} & & & & &    \\
\hline
\\
\endhead
E(MeV)     & 0.1   & 0.2           & 0.3       & 0.4          & 0.5          & 0.7          & 1           &  & 1           & 2           & 3           & 5           &  7 & 10  \\
\hline
5p1 &\ 1.15$_{  }$ & 1.11$_{  }$  & 1.07$_{  }$ & 1.05$_{  }$  & 1.03$_{  }$ & 9.99$_{-1}$  & 9.68$_{-1}$ &  & 9.72$_{-1}$ & 9.13$_{-1}$ & 8.84$_{-1}$ & 8.52$_{-1}$ & 8.36$_{-1}$ & 8.21$_{-1}$ \\
5p0 &\ 1.62$_{  }$ & 1.55$_{  }$  & 1.49$_{  }$ & 1.44$_{  }$  & 1.40$_{  }$ & 1.35$_{  }$  & 1.30$_{  }$ &  & 1.31$_{  }$ & 1.22$_{  }$ & 1.18$_{  }$ & 1.15$_{  }$ & 1.12$_{  }$ & 1.10$_{  }$ \\
5s  &\ 1.48$_{  }$ & 1.51$_{  }$  & 1.55$_{  }$ & 1.59$_{  }$  & 1.62$_{  }$ & 1.65$_{  }$  & 1.68$_{  }$ &  & 1.68$_{  }$ & 1.70$_{  }$ & 1.71$_{  }$ & 1.70$_{  }$ & 1.70$_{  }$ & 1.69$_{  }$ \\
4d2 &\ 1.29$_{  }$ & 1.14$_{  }$  & 1.08$_{  }$ & 1.03$_{  }$  & 1.00$_{  }$ & 9.60$_{-1}$  & 9.18$_{-1}$ &  & 9.22$_{-1}$ & 8.46$_{-1}$ & 8.04$_{-1}$ & 7.54$_{-1}$ & 7.22$_{-1}$ & 6.90$_{-1}$ \\
4d1 &\ 1.83$_{  }$ & 1.56$_{  }$  & 1.41$_{  }$ & 1.32$_{  }$  & 1.25$_{  }$ & 1.17$_{  }$  & 1.09$_{  }$ &  & 1.08$_{  }$ & 9.75$_{-1}$ & 9.20$_{-1}$ & 8.56$_{-1}$ & 8.16$_{-1}$ & 7.76$_{-1}$ \\
4d0 &\ 2.35$_{  }$ & 1.97$_{  }$  & 1.78$_{  }$ & 1.64$_{  }$  & 1.54$_{  }$ & 1.40$_{  }$  & 1.28$_{  }$ &  & 1.25$_{  }$ & 1.12$_{  }$ & 1.06$_{  }$ & 9.83$_{-1}$ & 9.33$_{-1}$ & 8.82$_{-1}$ \\
4p1 &\ 1.64$_{  }$ & 1.51$_{  }$  & 1.41$_{  }$ & 1.35$_{  }$  & 1.32$_{  }$ & 1.30$_{  }$  & 1.31$_{  }$ &  & 1.32$_{  }$ & 1.35$_{  }$ & 1.36$_{  }$ & 1.36$_{  }$ & 1.35$_{  }$ & 1.34$_{  }$ \\
4p0 &\ 3.02$_{  }$ & 2.44$_{  }$  & 2.23$_{  }$ & 2.12$_{  }$  & 2.05$_{  }$ & 2.00$_{  }$  & 2.05$_{  }$ &  & 2.02$_{  }$ & 2.06$_{  }$ & 2.04$_{  }$ & 2.00$_{  }$ & 1.98$_{  }$ & 1.95$_{  }$ \\
4s  &\ 1.98$_{  }$ & 1.79$_{  }$  & 1.71$_{  }$ & 1.69$_{  }$  & 1.69$_{  }$ & 1.73$_{  }$  & 1.73$_{  }$ &  & 1.65$_{  }$ & 1.63$_{  }$ & 1.64$_{  }$ & 1.62$_{  }$ & 1.60$_{  }$ & 1.57$_{  }$ \\
\end{longtable}
\newpage

%
%


\renewcommand{\arraystretch}{1.0}

\footnotesize

\begin{longtable}{@{\extracolsep\fill}lllllllllllllll@{}}
\caption{Mean impact parameter, $\left\langle
b_{nlm}^{1}\right\rangle$, for proton in Ne, normalized to the mean
radio of each subshell, $\left\langle r_{nl}\right\rangle $. See
page \pageref{tbl14te} for explanation.} \label{table14}
\mbox{Ne}  & \mbox{CDW-EIS} & & & & & &  & \ \ \ &\mbox{Born} & & & & &    \\
\hline
\\
\endhead
E(MeV)     & 0.1  & 0.2         & 0.3         & 0.4         & 0.5         & 0.7         & 1           & & 1           & 2           & 3           & 5           &  7 & 10  \\
\hline
2p1 & 1.85$_{  }$ & 1.91$_{  }$ & 2.00$_{  }$ & 2.09$_{  }$ & 2.17$_{  }$ & 2.33$_{  }$ & 2.53$_{  }$ & &  2.53$_{  }$ & 3.06$_{  }$ & 3.44$_{  }$ & 4.03$_{  }$& 4.49$_{  }$ & 5.05$_{  }$  \\
2p0 & 1.42$_{  }$ & 1.58$_{  }$ & 1.69$_{  }$ & 1.80$_{  }$ & 1.90$_{  }$ & 2.08$_{  }$ & 2.31$_{  }$ & &  2.33$_{  }$ & 2.92$_{  }$ & 3.35$_{  }$ & 4.01$_{  }$& 4.51$_{  }$ & 5.11$_{  }$  \\
2s  & 1.30$_{  }$ & 1.34$_{  }$ & 1.35$_{  }$ & 1.36$_{  }$ & 1.38$_{  }$ & 1.41$_{  }$ & 1.45$_{  }$ & &  1.45$_{  }$ & 1.59$_{  }$ & 1.70$_{  }$ & 1.87$_{  }$& 2.01$_{  }$ & 2.19$_{  }$  \\
1s  & 6.02$_{-1}$ & 7.33$_{-1}$ & 8.24$_{-1}$ & 9.00$_{-1}$ & 9.60$_{-1}$ & 1.06$_{  }$ & 1.18$_{  }$ & &  1.18$_{  }$ & 1.37$_{  }$ & 1.47$_{  }$ & 1.62$_{  }$& 1.73$_{  }$ & 1.86$_{  }$  \\
\end{longtable}
\bigskip

\renewcommand{\arraystretch}{1.0}

\footnotesize

\begin{longtable}{@{\extracolsep\fill}llllllllllllllr@{}}
\caption{Mean impact parameter, $\left\langle
b_{nlm}^{1}\right\rangle$, for proton in Ar, normalized to the mean
radio of each subshell, $\left\langle r_{nl}\right\rangle $. See
page \pageref{tbl15te} for explanation.} \label{table15}
\mbox{Ar}  & \mbox{CDW-EIS} & & & & & &  & \ \ \ &\mbox{Born} & & & & &    \\
\hline
\\
\endhead
E(MeV)     & 0.1   & 0.2           & 0.3       & 0.4          & 0.5          & 0.7          & 1         & &  1           & 2           & 3           & 5           &  7 & 10\ \ \ \   \\
\hline
3p1 & 1.55$_{  }$ & 1.67$_{  }$ & 1.79$_{  }$ & 1.90$_{  }$ & 2.00$_{  }$ & 2.16$_{  }$ & 2.36$_{  }$  & &  2.35$_{  }$   & 2.82$_{  }$  & 3.06$_{  }$ & 3.46$_{  }$ &   3.68$_{  }$  & 3.88$_{  }$\ \\
3p0 & 1.28$_{  }$ & 1.44$_{  }$ & 1.58$_{  }$ & 1.70$_{  }$ & 1.81$_{  }$ & 2.00$_{  }$ & 2.23$_{  }$  & &  2.22$_{  }$   & 2.74$_{  }$  & 3.01$_{  }$ & 3.44$_{  }$ &   3.66$_{  }$  & 3.87$_{  }$\ \\
3s  & 1.13$_{  }$ & 1.10$_{  }$ & 1.08$_{  }$ & 1.08$_{  }$ & 1.08$_{  }$ & 1.09$_{  }$ & 1.11$_{  }$  & &  1.10$_{  }$   & 1.17$_{  }$  & 1.24$_{  }$ & 1.34$_{  }$ &   1.42$_{  }$  & 1.53$_{  }$\ \\
2p1 & 9.38$_{-1}$ & 1.09$_{  }$ & 1.18$_{  }$ & 1.25$_{  }$ & 1.30$_{  }$ & 1.36$_{  }$ & 1.43$_{  }$  & &  1.42$_{  }$   & 1.58$_{  }$  & 1.70$_{  }$ & 1.90$_{  }$ &   2.05$_{  }$  & 2.25$_{  }$\ \\
2p0 & 7.55$_{-1}$ & 7.62$_{-1}$ & 8.01$_{-1}$ & 8.43$_{-1}$ & 8.88$_{-1}$ & 9.67$_{-1}$ & 1.07$_{  }$  & &  1.08$_{  }$   & 1.30$_{  }$  & 1.45$_{  }$ & 1.68$_{  }$ &   1.85$_{  }$  & 2.07$_{  }$\ \\
2s  & 6.96$_{-1}$ & 8.39$_{-1}$ & 9.50$_{-1}$ & 1.03$_{  }$ & 1.10$_{  }$ & 1.11$_{  }$ & 1.23$_{  }$  & &  1.21$_{  }$   & 1.28$_{  }$  & 1.32$_{  }$ & 1.40$_{  }$ &   1.47$_{  }$  & 1.55$_{  }$\ \\
1s  & 3.29$_{-1}$ & 4.40$_{-1}$ & 5.20$_{-1}$ & 5.83$_{-1}$ & 6.31$_{-1}$ & 7.02$_{-1}$ & 7.83$_{-1}$  & &  7.82$_{-1}$   & 9.69$_{-1}$  & 1.08$_{  }$ & 1.22$_{  }$ &   1.30$_{  }$  & 1.39$_{  }$\ \\
\end{longtable}
\bigskip

\renewcommand{\arraystretch}{1.0}

\footnotesize

\begin{longtable}{@{\extracolsep\fill}llllllllllllllr@{}}
\caption{Mean impact parameter, $\left\langle
b_{nlm}^{1}\right\rangle$, for proton in Kr, normalized to the mean
radio of each subshell, $\left\langle r_{nl}\right\rangle $. See
page \pageref{tbl16te} for explanation.} \label{table16}
\mbox{Kr} & \mbox{CDW-EIS} & & & & & &  & \ \ \ &\mbox{Born} & & & & &    \\
\hline
\\
\endhead
E(MeV)     & 0.1    & 0.2         & 0.3         & 0.4         & 0.5         & 0.7         & 1          & & 1           & 2           & 3           & 5           &  7 & 10\ \ \ \   \\
\hline
4p1 & 1.38$_{  }$  & 1.48$_{  }$ & 1.58$_{  }$ & 1.66$_{  }$ & 1.73$_{  }$ & 1.86$_{  }$ & 2.01$_{  }$& & 2.03$_{  }$ & 2.44$_{  }$ & 2.74$_{  }$ & 3.18$_{  }$  & 3.49$_{  }$  & 3.84$_{  }$ \\
4p0 & 1.12$_{  }$  & 1.27$_{  }$ & 1.38$_{  }$ & 1.48$_{  }$ & 1.56$_{  }$ & 1.71$_{  }$ & 1.88$_{  }$& & 1.91$_{  }$ & 2.36$_{  }$ & 2.69$_{  }$ & 3.18$_{  }$  & 3.53$_{  }$  & 3.90$_{  }$ \\
4s  & 1.10$_{  }$  & 1.07$_{  }$ & 1.05$_{  }$ & 1.04$_{  }$ & 1.04$_{  }$ & 1.04$_{  }$ & 1.04$_{  }$& & 1.05$_{  }$ & 1.07$_{  }$ & 1.09$_{  }$ & 1.12$_{  }$  & 1.16$_{  }$  & 1.20$_{  }$ \\
3d2 & 1.15$_{  }$  & 1.31$_{  }$ & 1.38$_{  }$ & 1.43$_{  }$ & 1.47$_{  }$ & 1.53$_{  }$ & 1.60$_{  }$& & 1.64$_{  }$ & 1.85$_{  }$ & 2.00$_{  }$ & 2.25$_{  }$  & 2.46$_{  }$  & 2.70$_{  }$ \\
3d1 & 9.59$_{-1}$  & 1.02$_{  }$ & 1.09$_{  }$ & 1.16$_{  }$ & 1.21$_{  }$ & 1.29$_{  }$ & 1.39$_{  }$& & 1.44$_{  }$ & 1.68$_{  }$ & 1.86$_{  }$ & 2.15$_{  }$  & 2.37$_{  }$  & 2.64$_{  }$ \\
3d0 & 8.46$_{-1}$  & 8.86$_{-1}$ & 9.41$_{-1}$ & 9.99$_{-1}$ & 1.05$_{  }$ & 1.15$_{  }$ & 1.26$_{  }$& & 1.34$_{  }$ & 1.62$_{  }$ & 1.82$_{  }$ & 2.12$_{  }$  & 2.35$_{  }$  & 2.63$_{  }$ \\
3p1 & 8.33$_{-1}$  & 9.85$_{-1}$ & 1.09$_{  }$ & 1.16$_{  }$ & 1.21$_{  }$ & 1.26$_{  }$ & 1.29$_{  }$& & 1.46$_{  }$ & 1.54$_{  }$ & 1.58$_{  }$ & 1.65$_{  }$  & 1.71$_{  }$  & 1.80$_{  }$ \\
3p0 & 6.43$_{-1}$  & 7.03$_{-1}$ & 7.41$_{-1}$ & 7.78$_{-1}$ & 8.18$_{-1}$ & 8.90$_{-1}$ & 9.64$_{-1}$& & 1.19$_{  }$ & 1.33$_{  }$ & 1.42$_{  }$ & 1.55$_{  }$  & 1.64$_{  }$  & 1.75$_{  }$ \\
3s  & 6.69$_{-1}$  & 7.99$_{-1}$ & 9.01$_{-1}$ & 9.80$_{-1}$ & 1.03$_{  }$ & 1.09$_{  }$ & 1.11$_{  }$& & 1.09$_{  }$ & 1.13$_{  }$ & 1.16$_{  }$ & 1.21$_{  }$  & 1.26$_{  }$  & 1.33$_{  }$ \\
\end{longtable}
\bigskip

\renewcommand{\arraystretch}{1.0}

\footnotesize

\begin{longtable}{@{\extracolsep\fill}llllllllllllllr@{}}
\caption{Mean impact parameter, $\left\langle
b_{nlm}^{1}\right\rangle$, for proton in Xe, normalized to the mean
radio of each subshell, $\left\langle r_{nl}\right\rangle $. See
page \pageref{tbl17te} for explanation.} \label{table17}
\mbox{Xe}  & \mbox{CDW-EIS} & & & & & &  & \ \ \ &\mbox{Born} & & & & &    \\
\hline
\\
\endhead
E(MeV)     & 0.1   & 0.2           & 0.3       & 0.4          & 0.5          & 0.7          & 1          & & 1           & 2           & 3           & 5           &  7 & 10\ \ \ \   \\
\hline
5p1 & 1.33$_{  }$ & 1.44$_{  }$  & 1.53$_{  }$ & 1.62$_{  }$  & 1.69$_{  }$ & 1.81$_{  }$ & 1.95$_{  }$ & & 1.94$_{  }$ & 2.25$_{  }$ & 2.44$_{  }$ & 2.65$_{  }$ & 2.77$_{  }$ & 2.89$_{  }$ \\
5p0 & 1.09$_{  }$ & 1.22$_{  }$  & 1.33$_{  }$ & 1.43$_{  }$  & 1.51$_{  }$ & 1.65$_{  }$ & 1.81$_{  }$ & & 1.81$_{  }$ & 2.15$_{  }$ & 2.35$_{  }$ & 2.57$_{  }$ & 2.70$_{  }$ & 2.81$_{  }$ \\
5s  & 1.03$_{  }$ & 1.02$_{  }$  & 1.01$_{  }$ & 1.00$_{  }$  & 1.00$_{  }$ & 9.99$_{-1}$ & 9.99$_{-1}$ & & 9.93$_{-1}$ & 1.01$_{  }$ & 1.02$_{  }$ & 1.04$_{  }$ & 1.06$_{  }$ & 1.08$_{  }$ \\
4d2 & 1.17$_{  }$ & 1.32$_{  }$  & 1.40$_{  }$ & 1.47$_{  }$  & 1.53$_{  }$ & 1.62$_{  }$ & 1.74$_{  }$ & & 1.73$_{  }$ & 2.03$_{  }$ & 2.26$_{  }$ & 2.61$_{  }$ & 2.88$_{  }$ & 3.20$_{  }$ \\
4d1 & 9.14$_{-1}$ & 1.06$_{  }$  & 1.17$_{  }$ & 1.26$_{  }$  & 1.33$_{  }$ & 1.45$_{  }$ & 1.59$_{  }$ & & 1.58$_{  }$ & 1.92$_{  }$ & 2.16$_{  }$ & 2.54$_{  }$ & 2.83$_{  }$ & 3.19$_{  }$ \\
4d0 & 8.01$_{-1}$ & 9.48$_{-1}$  & 1.05$_{  }$ & 1.14$_{  }$  & 1.22$_{  }$ & 1.35$_{  }$ & 1.50$_{  }$ & & 1.50$_{  }$ & 1.86$_{  }$ & 2.12$_{  }$ & 2.51$_{  }$ & 2.81$_{  }$ & 3.18$_{  }$ \\
4p1 & 8.20$_{-1}$ & 9.22$_{-1}$  & 1.01$_{  }$ & 1.06$_{  }$  & 1.09$_{  }$ & 1.11$_{  }$ & 1.12$_{  }$ & & 1.11$_{  }$ & 1.14$_{  }$ & 1.16$_{  }$ & 1.20$_{  }$ & 1.23$_{  }$ & 1.29$_{  }$ \\
4p0 & 5.81$_{-1}$ & 6.80$_{-1}$  & 7.27$_{-1}$ & 7.64$_{-1}$  & 7.93$_{-1}$ & 8.32$_{-1}$ & 8.48$_{-1}$ & & 8.52$_{-1}$ & 8.77$_{-1}$ & 9.02$_{-1}$ & 9.55$_{-1}$ & 1.00$_{  }$ & 1.07$_{  }$ \\
4s  & 7.11$_{-1}$ & 7.86$_{-1}$  & 8.54$_{-1}$ & 9.02$_{-1}$  & 9.27$_{-1}$ & 9.42$_{-1}$ & 9.49$_{-1}$ & & 9.64$_{-1}$ & 9.91$_{-1}$ & 1.02$_{  }$ & 1.07$_{  }$ & 1.11$_{  }$ & 1.17$_{  }$ \\
\end{longtable}

\end{document}